\documentclass[graybox]{svmult}

\usepackage{type1cm}        
\usepackage{comment} 
\usepackage{makeidx}         
\usepackage{graphicx}        
\usepackage{mathrsfs}
\usepackage{amsfonts}
\usepackage{nccmath}
\usepackage{multicol}        
\usepackage[bottom]{footmisc}

\usepackage{newtxtext}       %
\usepackage[varvw]{newtxmath}       

\usepackage{amssymb,mathtools}
\usepackage{amsmath}
\usepackage{subcaption}
\usepackage{booktabs,multirow}
\usepackage{algorithm,url,color}
\usepackage{algpseudocode}
\usepackage{algorithmicx}

\newcommand{\vect}[1]{\boldsymbol{\mathbf{#1}}}
\DeclareMathAlphabet{\mathpzc}{OT1}{pzc}{m}{it}

\algnewcommand{\LineComment}[1]{\State \(\#\) #1}
\algnewcommand\algorithmicinput{\textbf{Set}}
\algnewcommand\Set{\item[\algorithmicinput]}
\algnewcommand\algorithmicinitial{\textbf{Initialize}}
\algnewcommand\Initialize{\item[\algorithmicinitial]}

\let\oldReturn\Return
\renewcommand{\Return}{\State\oldReturn}

\makeindex             

\begin{document}

\title*{RIS-Enabled Smart Wireless Environments: Fundamentals and Distributed Optimization}
\author{George C. Alexandropoulos\orcidID{0000-0002-6587-1371} and\\ Kostantinos D. Katsanos\orcidID{0000-0003-1894-5216} and\\ George Stamatelis\orcidID{0000-0001-7826-2412} and\\ Ioannis Gavras\orcidID{0009-0007-8759-5930}}
\institute{George C. Alexandropoulos \at Department of Informatics and Telecommunications, National and Kapodistrian University of Athens,  Greece, \email{alexandg@di.uoa.gr}
\and Kostantinos D. Katsanos \at Department of Informatics and Telecommunications, National and Kapodistrian University of Athens,  Greece, \email{kkatsan@di.uoa.gr}
\and George Stamatelis \at Department of Informatics and Telecommunications, National and Kapodistrian University of Athens,  Greece, \email{georgestamat@di.uoa.gr}
\and Ioannis Gavras \at Department of Informatics and Telecommunications, National and Kapodistrian University of Athens,  Greece, \email{giannisgav@di.uoa.gr}}
%
%
\maketitle

\abstract{This chapter overviews the concept of Smart Wireless Environments (SWEs), initially motivated by the emerging technology of Reconfigurable Intelligent Surfaces (RISs). The operating principles and state-of-the-art hardware architectures of programmable metasurfaces are first introduced, encompassing passive, active, simultaneously reflecting and absorbing or transmitting, as well as non-local (a.k.a. Beyond-Diagonal (BD)) RIS designs, with the latter enabling additionally dynamic coupling among the metasurface's constituting unit elements. Subsequently, the chapter discusses key performance objectives and use cases of RIS-enabled SWEs, including spectral and energy efficiency, electromagnetic field exposure reduction and sustainability, reliability, physical-layer security, energy harvesting, localization/sensing and integrated sensing and communications, as well as the emerging paradigm of over-the-air computing. 
Focusing on the recent trend of BD-RISs, two distributed designs of respective SWEs are presented. The first deals with a multi-user Multiple-Input Single-Output (MISO) system operating within the area of influence of a SWE comprising multiple BD-RISs. A hybrid distributed and fusion machine learning framework based on multi-branch attention-based convolutional Neural Networks (NNs), NN parameter sharing, and neuroevolutionary training is presented, which enables online mapping of channel realizations to the configurations of the multiple distributed BD-RISs as well as the multi-user transmit precoder. The presented performance evaluation results showcase that the distributedly optimized RIS-enabled SWE achieves near-optimal sum-rate performance with low online computational complexity. The second design focuses on the wideband interference MISO broadcast channel setting, where each base station exclusively controls one BD-RIS to serve its assigned group of users. Considering a transmission line model for each BD-RIS, a cooperative optimization framework that jointly designs the base station transmit precoders as well as the tunable capacitances and switch matrices of all metasurfaces is presented. Numerical results demonstrating the superior sum-rate performance of the designed RIS-enabled SWE for multi-cell MISO networks over benchmark schemes, considering non-cooperative configuration and conventional diagonal metasurfaces, are presented. 
}

\section{Introduction}
\label{sec:intro_RIS}
The evolution from the $4$-th Generation (4G) to the $5$-th Generation (5G) wireless networks has been driven by the ever-increasing demand for higher data rates, massive connectivity, and stringent latency requirements. Despite substantial advances in radio interface design, network densification, and multi-antenna processing, current systems still operate within signal propagation environments that are essentially uncontrollable. In particular, the wireless channel is typically treated as a random, time-varying entity that must be estimated and compensated for, but cannot be shaped in a deterministic and application-aware manner. As a result, conventional network architectures largely rely on sophisticated transceiver-side processing, moderate-to-high transmit powers, and dense infrastructure deployments to counteract fading, blockages, and interference.

Reconfigurable Intelligent Surfaces (RISs), also referred to as intelligent reflecting surfaces, programmable metasurfaces, or large intelligent surfaces, have emerged as a key enabler for the paradigm of Smart Wireless Environments (SWEs)~\cite{Huang2019_RIS_EE,DiRenzo2019_SRE,alexandropoulos2020ris,9627818,Alexandropoulos2023_RISDeployment,cao2022massive,Liu2021_RIS_Survey,Jian2022_RIS_Hardware,Basar2024_RIS_6G}. By embedding electronically tunable metasurfaces into walls, ceilings, building facades, or other objects inside the signal propagation environment, the RIS technology allows the radio propagation medium to be dynamically programmed. In this vision, the wireless environment becomes a controllable asset~\cite{9482474} that can be jointly optimized with the transceivers to meet communication, sensing, energy transfer, or even lately over-the-air computing~\cite{Yang2023NextGenerationReconfigurableMetasurfaces,MINN}, objectives in a holistic way.

This chapter overviews the role of RISs as a pivotal technology for SWEs. First, the basic operating principles of programmable metasurfaces are presented, which are then followed by the latest advances in their hardware architectures. In the sequel, key performance metrics and optimization objectives that have largely driven the design of RIS-assisted wireless systems in the literature are described, including spectral and energy efficiency, ElectroMagnetic (EM) Field (EMF) exposure reduction, physical-layer security, energy harvesting, localization/sensing and Integrated Sensing And Communications (ISAC), as well as the emerging paradigm of Over-The-Air (OTA) computing. Representative deployment scenarios of RISs and the way they support the SWE paradigm are also discussed, with particular emphasis given on use cases targeting connectivity enhancement, localization, sensing, sustainability, and secrecy provisioning. 
Building on this foundation and considering the emerging hardware architecture of non-local (a.k.a. Beyond-Diagonal (BD))~\cite{Chen2025NonlocalMetamaterialsAndMetasurfaces,Demir2023_BDRIS}, which, apart from consisting of unit elements of tunable responses over the impinging signals, they enable dynamic element coupling, two representative distributed designs of respective SWEs are detailed and their performance is numerically evaluated and compared against relevant state-of-the-art benchmarks.

The remainder of this chapter is organized as follows. Section~\ref{subsec:what_is_ris} overviews the fundamental aspects of the RIS technology, while representative state-of-the-art design  approaches for RIS-enabled SWEs are discussed in Section~\ref{subsec:metrics_objectives}. Sections~\ref{sec:BD_RIS_NNs} and~\ref{Sec:IBC_RISs} present distributed optimization approaches for respectively single- and multi-cell Multiple-Input Single-Output (MISO) SWEs with multiple User Equipment (UE). The concluding remarks of the chapter are included in Section~\ref{sec:conclusions}. To facilitate a consistent exposition, the main notation adopted throughout the chapter is summarized in Table~\ref{table:notations}.
\begin{table}[!t]
  \centering
  \caption{The Notations of this Chapter.}
  \begin{tabular}{r|l}
    \toprule
    $a, \mathbf{a}$, and $\mathbf{A}$ &  Fonts for scalars, vectors, and matrices \\
    $\jmath \triangleq \sqrt{-1}$ & Imaginary unit \\
    $\Re\{\cdot\}$ & Real part of a complex scalar/vector/matrix\\
    $\Im\{\cdot\}$ & Imaginary part of a complex scalar/vector/matrix \\
    $\mathbf{A}^{\rm T}$ and $\mathbf{A}^{\rm H}$ &  Matrix transpose and Hermitian transpose \\
    $\mathbf{A}^{*}$ and $\mathbf{A}^{-1}$ &  Matrix conjugate and inverse \\ 
    $[\mathbf{A}]_{i,j}$ &  Matrix element at the $i$-th row and $j$-th column \\
    $[\mathbf{a}]_{i}$ &  The $i$-th vector element \\
    $\mathbf{I}_n$ & $n \times n$ ($n\geq2$) identity matrix \\
    $\mathbf{0}_{n \times k}$ & $n \times k$ matrix with zeros\\
    $\lvert \cdot \rvert$ & Amplitude of a complex scalar \\
    $\Vert \cdot \Vert$ & Vector Euclidean norm \\
    $\Vert \cdot \Vert_{\rm F}$ & Matrix Frobenius norm \\
    $\otimes$ & Kronecker product \\
    $\textrm{Tr}(\mathbf{A})$ & Trace of matrix $\mathbf{A}$ \\
    $\textrm{vec}(\mathbf{A})$ & Vectorization of $\mathbf{A}$ \\    
    $\textrm{vec}_{\rm d}(\mathbf{A})$ & Vector whose elements are the diagonal entries of the square matrix $\mathbf{A}$ \\   
    $\textrm{diag}(\mathbf{a})$ & Diagonal matrix with $\mathbf{a}$ on the main diagonal\\
    $\nabla_{\mathbf{a}} f(\mathbf{a})$ & Gradient vector of function $f(\mathbf{a})$ with respect to $\mathbf{a}$ \\
    $<\vect{A}_1,\vect{A}_2> \triangleq \Re\{\textrm{Tr}(\vect{A}_1^{\rm H}\vect{A}_2)\}$ & Inner product between matrices $\vect{A}_1$ and $\vect{A}_2$ of suitable dimensions \\
    ${\rm E}[\cdot]$ & Expectation operator \\
    ${\rm card}(\cdot)$ & Cardinality of a set\\
    $\mathbb{R}$ & Set of real numbers\\
    $\mathbb{C}$ & Set of complex numbers\\
    $\vect{x}\sim\mathcal{CN}(\vect{a},\vect{A})$ & Complex Gaussian random vector with mean $\vect{a}$ and covariance matrix $\vect{A}$ \\
    \bottomrule
  \end{tabular}
  \label{table:notations}
\end{table}

\section{Fundamentals of the RIS Technology}
\label{subsec:what_is_ris}
An RIS is a planar structure composed of a possibly extremely large number of low cost unit cells, also called meta-atoms or reflecting/unit elements, whose EM response can be electronically controlled~\cite{alexandropoulos2020ris}. Each element can independently adjust the phase and, in more advanced designs, the amplitude, polarization, or time delay of the incident EM wave~\cite{Jian2022_RIS_Hardware,Wu2021_IRS_Tutorial}. By jointly configuring all its unit elements, an RIS can shape the impinging wavefront in a programmable way, effectively implementing a controllable transformation of the wireless channel. From a signal processing perspective, a conventional wireless link between a Base Station (BS) and a UE is modeled by a channel matrix determined by the geometry of the environment, the EM properties of surrounding materials, and the distribution of scatterers~\cite{10453467}. When an RIS is deployed, an additional controllable channel propagation path is introduced comprising the BS-RIS and RIS-UE links. Together with the tunable RIS response profile, this path forms a RIS-parametrized channel whose effective characteristics can be reconfigured in real time. In this manner, the RIS operates as an almost passive, low power, and low noise environmental beamformer, which can complement or partially replace active relaying solutions and large-scale antenna deployments at the BS~\cite{Huang2019_RIS_EE}.

The initial and widely adopted consideration for the RIS hardware architecture comprises tunable, but purely reactive components (e.g., varactor diodes, PIN diodes, MEMS)~\cite{9475155}, which locally adjust the phase (and possibly amplitude) of the reflected signal without imposing any form of active Radio Frequency (RF) amplification. The power consumption of this architecture, which is loosely known as passive, is dominated by the biasing network and the control circuitry that are responsible for the reconfigurability of RIS unit elements, and is largely independent of the incident signal power~\cite{Alexandropoulos2023_RISDeployment}. Such almost passive RISs are thus particularly appealing for large-scale, low maintenance deployments, and are central in many of the studies discussed later in this chapter targeting energy efficiency and sustainability. 

However, the latter passive RISs suffer from a multiplicative pathloss effect, since the composite BS-RIS-UE channel experiences two signal propagation segments, and the metasurface does not provide any gain beyond coherent combining, when properly optimized. To mitigate this limitation, active~\cite{Tang2021_ActiveRIS,9758764,11045692} and hybrid~\cite{alexandropoulos2024hybrid} RIS hardware architectures have been proposed. In active RISs, some or all of their elements include power amplification stages, enabling partial compensation of the double pathloss, at the cost of higher hardware complexity and power consumption. For example, amplifying RIS designs with back-to-back metasurfaces and a shared power amplifier have been shown to significantly improve capacity and error-rate performance compared to purely passive counterparts~\cite{9758764}, while still consuming less power than conventional amplify-and-forward relays. Hybrid RIS hardware designs combine passive unit elements realizing simultaneous tunable reflection and absorption of the impinging signals~\cite{Alamzadeh2021_RIS_Sensing,birari2025dualdielectric}, or each of these operations in a time orthogonal manner~\cite{alexandropoulos2020hardware,R-RIS}, and possibly a small number of reception RF chains to enable collection of the absorbed signal in a baseband processor (usually installed at the RIS control unit) for further signal processing~\cite{10237986,HRIS_CE_all,he2023single}. 

Another important classification dimension of the RIS technology pertains to their actual operation mode. Early RIS implementations were solely reflecting, operating as programmable mirrors that redirect incident EM waves back towards the side from which they arrive. To overcome this inherent restriction, transmitting RISs~\cite{8743495} and Simultaneously Transmitting And Reflecting (STAR) RISs~\cite{9200683} have been proposed. Transmitting RISs operate as reconfigurable intelligent transmit arrays with fine-grained beamforming capabilities, forwarding energy to the opposite side of the surface with respect to the incident wave. STAR RISs further generalize this concept by enabling simultaneous and independently controllable reflection and transmission, thereby being capable of serving UEs located on both sides of the metasurface or offering UE positioning in the entire $360^\circ$ space~\cite{HeFakhreddineAlexandropoulos2023}. Closely related to transmitting and STAR configurations are lens-based RISs, or EM lenses, which are designed to implement spatially varying phase profiles that continuously focus/defocus the radiated energy, in a manner analogous to a passive lens.

Dynamic Metasurface Antennas (DMAs) constitute a closely related version of the programmable metasurface concept, relocating programmability from the wireless propagation environment to the transceiver aperture itself~\cite{shlezinger2021dma}. Whereas an RIS is typically deployed as an almost passive metasurface that reconfigures an existing channel through controllable reflection and/or transmission, a DMA is a reconfigurable radiating and/or absorbing panel populated by tunable metamaterial elements, whose EM response can be electronically programmed, and integrated on a panel with a feeding network (e.g., microstrips or two dimensional waveguides). This architecture enables compact, low cost, and energy efficient realizations of eXtremely Large (XL) apertures capable of performing hybrid digital and analog beamforming/combining, thereby serving as an enabler for the XL MIMO paradigm anticipated in $6$-th Generation (6G) BSs and access points~\cite{gavras2025nf_dma_thz_crb,gavriilidis2024nf_beamtracking_dma,gavras2025_dma_isac,gavras2025emc_dma_sensing,gavras2026_2Ddma_sensing}. From a signal processing and system modeling perspective, DMA arrays are often described through structured, hardware-constrained linear transformations applied to the transmitted and/or received signals. This abstraction is conceptually analogous to RIS parameterizations, yet acting within the RF transceiver hardware rather than on the external propagation channel.

Most early link-level models represent the RIS as a diagonal matrix operation over the wireless channel~\cite{Huang2019_RIS_EE,Wu2021_IRS_Tutorial,ChenHuAlexandropoulosGarciaArmada2022}, i.e., the constituent meta-atoms are assumed to operate independently and the overall surface response is described by a diagonal matrix whose entries are complex reflection coefficients. This abstraction is both analytically convenient and sufficiently accurate for a wide range of practical designs~\cite{10453467}, and it has therefore underpinned a substantial body of work on RIS-assisted wireless communication applications. To move beyond this independent-element assumption, recent research has investigated BD-RIS hardware architectures~\cite{Demir2023_BDRIS,Li2023_BDRIS_Wideband,Demir2022_WidebandBDRIS,wijekoon2025physically}, in which coupling networks are intentionally introduced among the unit elements. Such coupling enables the metasurface to realize more general linear transformations of the impinging EM field, so that each outgoing component may depend on multiple incoming components, with jointly adjustable amplitude and phase. By providing additional degrees of freedom for wavefront shaping, BD-RISs can, in principle, offer improved capacity and interference mitigation relative to diagonal RISs, albeit at the expense of increased hardware complexity. Finally, it is worth noting that the RIS concept extends naturally to millimeter-Wave (mmWave) and sub-THz systems, where recent measurement campaigns of novel RIS hardware designs have begun to quantify RIS-induced channel characteristics in indoor environments~\cite{10500968,alexandropoulos2025characterization}.
 
\section{State-of-the-Art Designs for RIS-Enabled SWEs}
\label{subsec:metrics_objectives}
The literature on RIS-empowered wireless systems investigates diverse optimization criteria, including connectivity enhancement and reliability, energy efficiency and sustainability, physical-layer security, energy harvesting, localization/sensing and ISAC, as well as OTA computing. This section provides a high-level overview of the principal metrics considered in state-of-the-art RIS studies and summarizes the approaches that programmable metasurfaces are typically designed to optimize them.

\subsection{Communication Metrics}
A primary objective in many designs for RIS-assisted wireless systems is the enhancement of the Spectral Efficiency (SE) metric, which quantifies the mutual information between the transmitted and received signals. In this context, SE maximization problem formulations commonly require the joint design of the transmit beamforming vectors at one or multiple BSs, the configuration of one or more RISs (i.e., responses of their unit elements and also the inter-element coupling in the case of BD-RISs), and, when applicable, auxiliary resource-allocation variables, such as power control, user scheduling, and the adopted multiple access strategy. Fundamental analyses have characterized the mean and variance of the mutual information in multi-antenna systems assisted by multiple RISs, and have demonstrated that suitably optimized metasurfaces can yield substantial SE gains, particularly in propagation conditions with limited angular spread~\cite{Moustakas_Cap2023}. Alongside optimization-based formulations, RIS control schemes driven by Machine Learning (ML) methods have been investigated for rate-oriented objectives, including indoor signal focusing via learned phase configurations~\cite{huang2019indoor}. Complementary works have investigated the capacity and achievable rate regions of multi-user deployments in which a BS serves multiple UEs via an RIS~\cite{moustakas2024mimo}, including comparisons between Non-Orthogonal Multiple Access (NOMA) and orthogonal counterparts. Across both single- and multi-RIS settings, the RIS configuration and, crucially, its spatial placement have been shown to markedly affect the achievable sum rate and user fairness, exhibiting pronounced benefits in challenging regimes such as blockages and cell-edge operation~\cite{Wu2019_IRS_Beamforming,Bjornson2020_CapacityRIS}. For instance, RISs placed
near cell boundaries can extend coverage, reduce the frequency of handovers, mitigate
inter-cell interference, and improve connection reliability. In severely
obstructed environments, communications may even rely predominantly on RIS-mediated links, whose coordinated configuration resembles multi-cell
coordinated beamforming~\cite{AlexandropoulosFerrandGorcePapadias2016}, while additionally enabling direct control of the radio propagation environment~\cite{huang2021multihop}.

Energy Efficiency (EE) constitutes a central performance metric in RIS-enabled SWEs, reflecting the drive towards energy-sustainable network operation~\cite{Huang2019_RIS_EE}. This metric quantifies the ratio between the achievable sum-rate performance and the total consumed power, usually accounting for both radiated power and power consumption attributed to hardware-related circuitry.
EE-oriented designs often lead to non-convex optimization problems in which BS beamforming and RIS phase configurations (and, in some settings, the number and placement of RISs) are jointly optimized under quality-of-service and/or power constraints. Studies on downlink multi-UE MISO systems have shown that appropriately configured RISs can deliver substantial EE improvements relative to conventional amplify-and-forward relays, even in regimes where those relays may attain higher SE. Similar conclusions extend to wideband and cell-free deployments with multiple BSs and multiple RISs~\cite{katsanos2022wideband_all,Katsanos_distributed_all,IBC_distributed2024_all,Katsanos_Cell_Free_conf}, where redistributing functionality from power-hungry active infrastructure towards a larger number of low power metasurfaces can improve system-level EE, particularly when UEs are located in the vicinity of the RISs. Beyond classical EE, sustainability considerations in RIS-enabled SWEs also include EMF exposure control. Accordingly, EMF-aware utilities, such as self-EMF exposure utility (relating a UE's rate to its own exposure) and inter-EMF exposure utility (relating a target UE's rate to the maximum exposure experienced by surrounding UEs), have been proposed to capture trade-offs among throughput, power consumption, and EMF constraints~\cite{Alexandropoulos2023_RISDeployment,DT_EMF}. In this respect, RISs are particularly valuable devices because they enable radiation shaping that concentrates energy toward intended UEs while reducing unintended leakage to other spatial directions~\cite{alexandropoulos2020ris}.

The RIS technology has been also established as instrumental in enhancing link reliability, particularly in propagation-challenging scenarios. Reliability is commonly assessed via Outage Probability (OP), coverage probability, and bit error rate. By engineering an additional controllable propagation path, an RIS can mitigate deep fades and, in blockage-dominated deployments where the direct BS-UE channel is severely attenuated or unavailable, effectively restore connectivity and reduce the likelihood of outage. Accordingly, analytical studies have derived closed-form or tight approximations for the OP of RIS-assisted wireless links, and have quantified this metric's dependence on the number of RIS elements, the accuracy of phase alignment, and the fading statistics of the constituent BS-RIS and RIS-UE channels~\cite{Huang2019_RIS_EE,Wu2019_IRS_Beamforming,Selimis2021_NakagamiRIS,Lin2021_RIS_RPM}. 

In addition, RISs have been extensively studied as a means of strengthening secrecy performance, which is commonly quantified through the secrecy rate, defined as the positive part of the difference between the capacity of the legitimate link and that of the eavesdropping one. In fading environments, secrecy performance is often characterized by the secrecy OP, i.e., the probability that the instantaneous secrecy rate falls below a prescribed threshold, and by the average secrecy capacity, which captures the long-term attainable secrecy rate. By appropriately configuring the surface response, an RIS can reshape the propagation environment to promote constructive combining at the intended receiver while inducing destructive interference and/or spatial nulls towards potential eavesdroppers~\cite{safeguarding_ICC21}. Accordingly, secrecy-oriented studies have shown that increasing the size of a legitimate RIS, in terms of the number of controllable elements, can deliver substantial secrecy gains, whereas a malicious RIS under adversarial control may severely degrade secrecy, unless its effect is explicitly counteracted through robust designs of a legitimate programmable metasurface~\cite{Yang2020_RIS_Secrecy,Alexandropoulos2023_RIS_Secrecy,Katsanos_Spatial_2023}.

In general, the RIS configuration optimization is posed as a per-channel, per-slot design problem
under idealized Channel State Information (CSI) availability assumptions. In practical deployments, however, RIS configuration will need to be updated rapidly and at scale under UE mobility, imperfect and delayed CSI, as well as stringent constraints on training, signaling,
and control overhead~\cite{10600711}. These challenges are further exacerbated in multi-RIS and distributed system architectures, where the
dimensionality of the configuration space grows quickly and centralized,
iterative optimization becomes increasingly costly~\cite{RIS_Vahid_low}. This has motivated ML-driven configuration strategies that directly map
measured context (e.g., locations, angles, and channel features) to RIS settings with low online complexity, while shifting the computational burden to offline training and data collection~\cite{alexandropoulos2020phase,Alexandropoulos2021_PervasiveML,mbacnn}. Such approaches enable lightweight inference at run time, facilitate
distributed control, and offer improved robustness to model mismatch and CSI
uncertainty. In this direction, RIS-assisted networks are increasingly
interpreted through the lens of a SWE, where the propagation medium is treated as an additional, programmable resource and ML tools provide a principled means to support reliable operation across time-varying conditions and heterogeneous deployment constraints.

\subsection{Energy Harvesting}
In energy harvesting applications, performance is commonly quantified by the harvested power at energy receivers and by the Energy OP (EOP), with the latter measuring the probability that the harvested energy falls below a prescribed threshold. When an RIS is employed to steer and concentrate RF energy towards energy harvesting nodes, analytical tools developed for capacity and reliability evaluation can be readily adapted to obtain closed-form expressions, approximations, or bounds for the resulting EOP. A particularly relevant research direction concerns the energy autonomicity of the RIS structure itself~\cite{Ntontin2021_AutonomousRIS,Zhang2023_RIS_EH_Performance,10693440}. In self-sustainable (a.k.a. energy-autonomous) RIS hardware architectures and frameworks, a fraction of their unit elements may operate in an absorption mode, harvesting energy from incident information-bearing signals to empower the RIS control circuitry, rather than solely contributing to reflective beamforming. The associated design problems typically entail allocating unit elements between communications and energy harvesting functionalities, while jointly optimizing BS beamforming and RIS coefficients to ensure RIS self-sufficiency without compromising end-to-end communication quality. This, in turn, motivates a
closed-loop design between transceivers and programmable metasurfaces to jointly
optimize efficiency, autonomy, and reliability~\cite{masaracchia2024metaverse}. Related formulations also consider the coexistence of RIS-assisted communications and energy harvesting for dedicated external energy UEs, leading to coupled performance measures such as typical information OP for information UEs and EOP for energy users within a unified optimization framework.

\subsection{Localization, Sensing, and Integrated Sensing and Communications}
The RIS technology has been widely regarded as key enabler for localization and sensing objectives as well as for the ISAC paradigm~\cite{chepuri2023isacris,strinati2025distributed,Liu2023_RIS_ISAC_Overview}. In localization and sensing applications, performance is commonly assessed through position-error bounds (e.g., the Cram\'er--Rao lower bound and related Fisher-information metrics), achievable resolution in range/angle/Doppler, detection and false-alarm probabilities, and, in mapping tasks, the accuracy of reconstructed environmental features~\cite{kim2024mapping}. By introducing controllable interactions with the propagation medium, RISs can generate virtual Line-of-Sight (LoS) components via coordinated reflections in otherwise obstructed non-LoS regions, increase angular and delay diversity by creating additional specular paths, and improve geometric observability for position and orientation estimation, even in the absence of direct BS–UE links~\cite{RISLocalization2023_Feasibility}. The latter includes RIS-enabled SWE implementation for smart cities where sidelink-enabled operation can support seamless localization and sensing~\cite{chen2023smartcities}. Moreover, lens-type and focusing RIS configurations can implement spatially varying phase profiles that yield location-dependent field patterns~\cite{kim2024mapping,9827873,Rahal2023RISNearfield}, which can be exploited to enhance near-field localization and increase sensitivity to UE position variations. 

The latter capabilities have motivated RIS-assisted localization and sensing frameworks in which the metasurface is optimized to maximize localization accuracy or to improve detection performance, including joint designs that leverage both direct and RIS-reflected paths for tighter position error bounds and improved robustness under blockage~\cite{Alamzadeh2021_RIS_Sensing,Liu2023_RIS_ISAC_Overview,kim2023ris_monostatic_iccworkshops}, as well as mmWave-oriented approaches that jointly address channel acquisition and localization under practical hardware nonidealities~\cite{bayraktar2024dictionary}. Additionally, hybrid RIS hardware architectures, which combine largely passive meta-atoms with a small number of active sensing/processing interfaces~\cite{alexandropoulos2024hybrid,birari2025dualdielectric,R-RIS}, provide a practical means to endow the metasurface panel with situational awareness and to enable closed-loop configuration based on observed propagation and UE dynamics~\cite{gavras2025trackingaided_hybrid_ris}. This capability is particularly relevant for ISAC, as hybrid RISs can support simultaneous communication and sensing operations, also facilitating integrated designs that incorporate security constraints alongside sensing and communication objectives~\cite{gavras2025simultaneous_hybrid_ris,gavras2025secure_isac_hybrid_ris}. In full ISAC deployments, the RIS configuration needs to be selected to simultaneously support data transmission and sensing, thereby inducing inherent trade-offs between communication objectives (e.g., rate, reliability, and energy efficiency) and sensing objectives (e.g., localization accuracy and estimation resolution). Consequently, RIS-assisted ISAC typically leads to multi-objective or constrained optimization formulations in which BS precoding, RIS reflection coefficients, and resource allocation are jointly tuned to meet dual-functionality requirements~\cite{chepuri2023isacris,strinati2025distributed}.

\subsection{Over-the-Air Computation}
Beyond shaping wireless channels for communication- and sensing-centric objectives, programmable metasurfaces can also enable OTA computations (a.k.a. wave-domain computing)~\cite{alexandropoulos2020ris,9475155,wang2024_otac_6g_foundations}. This emerging computing paradigm envisions part of the transceiver signal/information processing tasks~\cite{AnYuenGuanDiRenzoDebbahPoorHanzo2024}, or others~\cite{stylianopoulos2025_minn_otainference,Kyriakos_icassp,Pandolfo2025}, to be effectively realized through appropriately optimized wave propagation enabled by controllable EM response of the wireless environment, thereby reducing the reliance on purely digital processing and/or analog/RF pipelines~\cite{Omam_WaveComputing}. For this purpose, RISs can be configured to sculpt the effective multi-UE channel so as to support reliable and accurate functional aggregation, and, more broadly, they can be interpreted as computational reconfigurable metasurfaces that tightly couple wave-domain transformations with task-oriented computing objectives \cite{yang2023_computational_ris,HuangAnYangGanBennisDebbah2024}. This perspective has been further advanced by metasurface-integrated deep Neural Network (NN) formulations~\cite{stylianopoulos2025_minn_otainference}, in which one or more programmable metasurfaces are optimized jointly with digital processing blocks in an end-to-end manner, enabling OTA edge inference with reduced communication overhead and enhanced robustness to channel impairments~\cite{stylianopoulos2025_minn_otainference}. Moreover, Stacked Intelligent Metasurfaces (SIM), comprising multiple layers of programmable metasurfaces, can realize richer and higher-dimensional wave-domain transformations than single-layer surfaces, and have been advocated as a scalable hardware platform for analog signal processing in next generation networks, e.g., holographic MIMO
\cite{an2023sim_holographic_mimo}. Collectively, RIS- and SIM-enabled OTA computation calls for joint designs spanning waveform and precoding/combining synthesis, metasurface configuration, and ML objectives, while also introducing practical challenges related to synchronization, channel acquisition, hardware constraints, and robustness to noise and model mismatch \cite{wang2024_otac_6g_foundations,yang2023_computational_ris}.

\section{Optimization of RIS-Enabled Broadcast SWEs}
\label{sec:BD_RIS_NNs}
In this section, a multi-UE Multiple-Input Single-Output (MISO) system operating inside the area of influence of a SWE comprising multiple BD-RISs is considered, with the design objective to jointly optimize the BS transmit precoder and the configurations of the SWE's multiple programmable metasurfaces for maximizing the achievable sum-rate performance. An NN architecture targeting the online optimization of the system's free parameters under realistic constraints for the BD-RISs and the overall BS precoding matrix is presented. The proposed NN is designed to be installed at the controller of each BD-RIS to decide, relying on the availability of locally relevant CSI, their individual configurations as well as a candidate set of BS precoders for all the UEs. The latter distributedly computed candidate precoding designs from all BD-RISs are then transferred and fused at the BS side to construct the final BS precoding matrix. Numerical results demonstrate that the proposed RIS-enabled SWE achieves near-optimal performance with affordable computational complexity.

\subsection{System and Received Signal Models}
The considered RIS-enabled MISO broadcast channel includes an $N_{\rm tx}$-antenna BS wishing to communicate with $N_{\rm ue}$ single-antenna UEs via the assistance of a SWE comprising $K$ identical BD-RISs. In various practical scenarios, the direct BS-UE links can be highly faded and the metasurfaces are used to enhance received signal quality. Wireless communication is structured over sequences of channel coherence blocks, indexed by $t$, according to which the channel gain is modeled as a random variable, remaining quasi-static within each block and changing independently and identically distributed from one block to the next one.

It is assumed that the RIS-empowered multi-UE MISO communication system lacks of a global fusion center responsible for deciding on the optimal configuration of all BD-RISs. Instead, the considered RIS-enabled SWE optimizes its individual BD-RIS in a distributed manner. In addition, the BS designs the precoding vectors $\mathbf{v}_1(t),\ldots,\mathbf{v}_{N_{\rm ue}}(t) \in \mathbb{C}^{N_{\rm tx} \times 1}$, with each intended for a specific UE and selected from a discrete codebook $\mathcal{V}\triangleq\{ [\mathbf{F}_{\rm DFT}]_{:,i}\}_{i=1}^{N_{\rm tx}}$, whose elements correspond to the columns of the $N_{\rm tx} \times N_{\rm tx}$ Discrete Fourier Transform (DFT) matrix $\mathbf{F}_{\rm DFT}$. It is noted that this set of precoders is a simplified version of the 3GPP 5G New Radio Type I codebook~\cite{3GPP_TS_38.214_codebooks}.

\subsubsection{BD-RIS Model}
Each BD-RIS structure consists of the metasurface panel hosting the response-tunable unit elements and a local controller~\cite{Basar2024_RIS_6G}, which determines dynamically the element responses and their interconnections via the proposed NN architecture. A batch BD-RIS model with ON/OFF connection switches signifying the inter-element coupling is considered~\cite{BD_trans_IT}, according to which the $N_{\rm ris}\times N_{\rm ris}$ matrix $\boldsymbol{\Phi}_k(t)$, including the configuration of the elements of each $k$-th ($k=1,\ldots,K$) metasurface for each $t$-th channel block, has nonzero entries only in the main diagonal and the $N_{\rm B}$ superdiagonals and subdiagonals. For example, $N_{\rm B}=1$ indicates that switches are only present between neighboring unit elements, whereas, for $N_{\rm B}=2$, connections are allowed between neighboring elements with distance $1$ and $2$ (i.e., for each $(i,j)$-th element pair with $|i-j| \leq 2$), and so on. It is noted that the optimal values of the BS precoders $\mathbf{v}_n(t)$'s ($n=1,\ldots,N_{\rm ue}$) will depend on all selected BD-RIS profiles, hence, $K$ adequate control links from the distinct controllers of the metasurfaces to the BS are assumed to be present~\cite{10600711,KunzBaskaranAlexandropoulos2025} to enable fusion of the respective outputs of the individual NNs. As will be described in the sequel, the controller of each $k$-th BD-RIS outputs a candidate set of BS precoding vectors $\mathbf{v}_{k,1}(t),\ldots,\mathbf{v}_{k,N_{\rm ue}}(t)$, and all sets are then used by the BS that deploys an intelligent fusion mechanism to select the final precoding matrix for all UEs; notation $\mathbf{V}(t)\in \mathbb{C}^{N_{\rm tx} \times N_{\rm ue}}$ will be used to denote the matrix with all selected vectors at each time instant $t$.

In order to be consistent with practical hardware implementations of RISs (including diagonal RISs, i.e., absent of mutual couplings among elements), it is assumed that all nonzero elements of each $\boldsymbol{\Phi}_k(t)$ are constrained in a discrete set~\cite{Alexandropoulos2023_RISDeployment,10500968,Alexandropoulos2021_PervasiveML,Stylianop1Bit,MIMOris1bit,stamatelisPowerControl,StamatelisRISDetection}. 
For notational convenience, the set of all applicable $N_{\rm B}$-batch BD-RIS configuration matrices, including the constraints for the feasible response states per unit element as well as the switch-based BD-RIS architecture, is represented by $\mathcal{D}_{\rm B}$.

\subsubsection{Received Signal Model}
The channel vectors and matrices involved in the considered RIS-empowered downlink MISO transmission are defined as follows:
\begin{itemize}
    \item $\mathbf{h}_n(t) \in \mathbb{C}^{N_{\rm tx} \times 1}$: The channel vector between the BS and each $n$-th UE.
    \item $\mathbf{H}_{1,k}(t) \in \mathbb{C}^{N_{\rm tx} \times N_{\rm ris}}$: The channel matrix between the BS and each $k$-th BD-RIS.
    \item $\mathbf{h}_{2,n,k}(t) \in \mathbb{C}^{N_{\rm ris} \times 1}$: The channel vector between each $k$-th BD-RIS and each $n$-th UE.
\end{itemize}
Assuming that signal reflections involving multiple BD-RISs (i.e., inter-metasurface paths) result in negligible power contributions and can be thus ignored \cite{Alexandropoulos2021_PervasiveML,mbacnn}, the effective combined channel vector $\mathbf{m}_n(t) \in \mathbb{C}^{1 \times N_{\rm tx}}$ from the BS to each $n$-th UE is given by the superposition of the direct path and the $K$ reflected paths via the BD-RISs as:
\begin{equation}
    \mathbf{m}_n(t) \triangleq \mathbf{h}_n^{\rm H}(t) + \sum_{k=1}^{K} \mathbf{h}_{2,n,k}^{\rm H}(t) \mathbf{\Phi}_k(t) \mathbf{H}_{1,k}^{\rm H}(t).
    \label{eq:effective_channel}
\end{equation}

Let ${x}_1(t),x_2(t),\ldots,x_{N_{\rm ue}}(t)$ denote the complex-valued transmitted data symbols each with power ${\rm E}[|x_n(t)|^2]=P$ ($N_{\rm ue}P$ denotes the total power budget for data transmission) at each time instant $t$, and $\tilde{n}(t)\sim\mathcal{CN}(0,\sigma^2)$ represent the Additive White Gaussian Noise (AWGN). The typical assumption that AWGN's variance $\sigma^2$ can be reliably estimated, and thus it can be perfectly known, is made. The baseband received signal at each $n$-th UE can therefore be expressed as follows: 
\begin{equation}
    y_n(t) \triangleq \underbrace{\mathbf{m}_n(t) \mathbf{v}_n(t) x_n(t)}_{\text{desired Signal}} + \underbrace{\sum_{j=1, j \neq n}^{N_{\rm ue}} \mathbf{m}_n(t) \mathbf{v}_j(t) x_j(t)}_{\text{multi-UE interference}} + \tilde{n}(t).
    \label{eq:received_signal}
\end{equation}

\subsection{System Design Objective}
The Signal-to-Interference-plus-Noise Ratio (SINR) at each $n$-th UE at each time instant $t$ is formulated as follows:
\begin{equation}
    \gamma_n(t) \triangleq \frac{|\mathbf{m}_n(t) \mathbf{v}_n(t)|^2}{\sum_{j=1, j \neq n}^{N_{\rm ue}} |\mathbf{m}_n(t) \mathbf{v}_j(t)|^2 + \frac{\sigma^2}{P}}.
    \label{eq:sinr}
\end{equation}
Consequently, the instantaneous achievable sum-rate performance per time instance $t$ can be computed as $R(t)\triangleq\sum_{n=1}^{N_{\rm ue}}R_n(t)$ with $R_n(t)\triangleq\log_2(1+\gamma_n(t))$ denoting the achievable rate for each $n$-th UE. 

Hereinafter, the design objective for the considered RIS-enabled broadcast SWE is the joint configuration of the BD-RIS profiles and the BS precoding vectors at each time instance $t$ that maximizes $R(t)$. It is assumed that, at each channel coherence block, all involved channel gains are estimated using established CSI techniques (see \cite{Jian2022_RIS_Hardware,HRIS_CE_all,RIS_Vahid_low,Swindlehurst_CE} and references therein), and then provided to the sum-rate optimization algorithm. Formally, by setting $\bar{\boldsymbol{\Phi}}(t)\triangleq\{\boldsymbol{\Phi}_1(t),\ldots,\boldsymbol{\Phi}_K(t)\}$, this design problem can be mathematically formulated as: 
\begin{equation*}
\begin{split}
  \mathcal{OP}_1: &\max_{\bar{\boldsymbol{\Phi}}(t),\mathbf{V}(t)} \quad R(t)
	\\& \textrm{s.t.}~~ ({\rm C1}): \boldsymbol{\Phi}_k(t) \in \mathcal{D}_{\rm B} \,\,\forall k,
	 \\&~~~\quad ({\rm C2}): \mathbf{v}_n(t) \in \mathcal{V}  \quad \forall n. 
\end{split}
\end{equation*}
Typically, due to the discrete nature of the constraints, the solution for this SWE design problem can be obtained via iterative discrete optimization approaches. However, discrete optimization is NP-hard, implying that the execution of the optimization algorithm needs to be completed well within the channel coherence time, which is expected to be in the order of few milliseconds. This attribute of real-time decision making is crucial in the foreseen RIS-enabled SWEs, since computation-induced latency may not only lead to outdated CSI at the time of configuration, but also to reduce SE due to the unavoidable idle time depriving data transmission. 

\subsubsection{$\mathcal{OP}_1$'s Relaxation via Centralized NN-Parametrized Mapping}
Let the following matrix definitions: 
\begin{align}
    \boldsymbol{\rm H}_{\rm D}(t) &\triangleq [\mathbf{h}_1(t),\ldots,\mathbf{h}_{N_{\rm ue}}(t)]\in \mathbb{C}^{N_{\rm tx} \times N_{\rm ue}}
    ,\\
    \boldsymbol{\rm H}_{\rm I,1}(t)&\triangleq[\mathbf{H}_{1,1}(t),\ldots,\mathbf{H}_{1,K}(t)] \in \mathbb{C}^{N_{\rm tx} \times KN_{\rm ris}},\\    
    \boldsymbol{\rm H}_{\rm I,2}(t)&\triangleq[\mathbf{h}_{2,1,1}(t),\mathbf{h}_{2,2,1}(t),\ldots,\mathbf{h}_{2,N_{\rm ue},K}(t)] \in \mathbb{C}^{N_{\rm ris} \times KN_{\rm ue}},
\end{align}
as well as the mapping/decision function $g(\cdot)$ (or policy) intended to satisfy $\mathcal{OP}_1$'s objective by mapping the instantaneous channel matrices to configurations of the $K$ BD-RISs and the BS precoding vectors, i.e.: 
\begin{equation}\label{eq:mapping-function}
    \{\bar{\boldsymbol{\Phi}}(t),\mathbf{V}(t)\} = g\left(
    \mathbf{H}_{\rm D}(t),\mathbf{H}_{\rm I,1}(t),\mathbf{H}_{\rm I,2}(t)
    \right).
\end{equation}
As previously mentioned, to address the computational limitations associated with iteratively solving $\mathcal{OP}_1$ at every time instance $t$, the computational requirements for this mapping function need to be small.  By using~\eqref{eq:mapping-function}, it is convenient to re-express the instantaneous achievable sum-rate performance, $R(t)$, as follows:
\begin{align}\label{eq:SNR-re-def}
R\left(\bar{\boldsymbol{\Phi}}(t), \mathbf{V}(t)\right) \triangleq R\left(g\left(\mathbf{H}_{\rm D}(t),\mathbf{H}_{\rm I,1}(t),\mathbf{H}_{\rm I,2}(t)\right)\right),
\end{align}
emphasizing the role of the mapping function $g(\cdot)$. To alleviate from the computational cost of per-time-instant optimization, it is proposed to relax $\mathcal{OP}_1$, focusing on finding a general mapping function that can perform online near-optimal decisions for a wide variety of channel inputs. 

Consider a finite time horizon $T$ so that $t=1,\ldots,T$.
At every time instant $t$, \eqref{eq:mapping-function} can be invoked to design the pair $\{ \bar{\boldsymbol{\Phi}}(t),\mathbf{V}(t)\}$ upon observing the current CSI. To this end, a sum-SNR maximization objective over the time horizon can be adopted which is expressed mathematically as follows:
\begin{equation*}
  \mathcal{OP}_2: \max_{g(\cdot) \in \mathcal{G}}  {\rm E}\left[ 
  \sum_{t=1}^{T} R\left(g\left(\mathbf{H}_{\rm D}(t),\mathbf{H}_{\rm I,1}(t),\mathbf{H}_{\rm I,2}(t)\right)\right) \right]~\hspace{0.2cm}\textrm{s.t.}~\hspace{0.2cm}({\rm C1})\,\,{\rm and}\,\,({\rm C2}),
\end{equation*}
where $\mathcal{G}$ represents the set of all applicable general mappings, i.e., it contains any function capable of mapping global CSI to the feasible sets of
the BD-RIS configurations and BS precoding matrices. In this formulation, the AWGN and the channel matrices are treated as random values, and the expectation is taken with respect to their joint probability density function over the $T$ time instants. Since the joint distributions involved are intractable, the expectation can be treated via optimizing $g(\cdot)$ over a sequence of Monte Carlo (MC) episodes of independent random value samples.

To find $g(\cdot)$ solving $OP_2$, an elaborate NN model can be used to parameterize this general mapping function~\cite{mbacnn}. For this purpose, the optimization of the network's weights can be conducted offline by simulating large numbers of UE trajectories of length $T$, and then, the expectation over the channels can be approximated by averaging the sums of all trajectories. In this way, during the online testing/deployment phase of the optimized NN and, at each channel coherence time instance $t$, the channels will be first estimated via existing CSI estimation methods, and then, provided to the NN which will quickly decide on the near-optimal pair of BD-RIS configurations $\bar{\boldsymbol{\Phi}}(t)$ and BS precoding matrix $\mathbf{V}(t)$. However, this centralized ML approach, necessitating the installation of a central processing unit (e.g., a central Graphics Processing Unit (GPU) at the BS), entails the following inherent limitations:
\begin{itemize}
    \item \textbf{Increased control overhead:} Instantaneous CSI of all involved channels needs to be available at each time instance $t$ at the central processing unit via dedicated control links of non-negligible capacity~\cite{10600711}. In fact, those control links need to support bi-directional control data communications~\cite{9627818} for the $\mathcal{OP}_2$-optimizing BD-RIS configuration matrices $\boldsymbol{\Phi}_k(t)$'s calculated by the central processor (at the BS side) to be transferred to each $k$-th BD-RIS controller. Evidently, this approach is associated with a large overhead in terms of both control signaling and synchronization. 
    \item \textbf{Large action and observation spaces:} The dimensions of both the input and output matrices exhibit a $K$-fold increase, implying that the mapping function between such high-dimensional spaces might be too complicated to efficiently approximate and optimize. Furthermore, even during inference time, the size of the model may be too computationally demanding to support online (i.e., real-time) control of the SWE.
\end{itemize}

\subsubsection{$\mathcal{OP}_1$'s Relaxation via Distributed NN-Parametrized Mapping}\label{sec:distributed}
Recent hybrid RIS frameworks~\cite{Alamzadeh2021_RIS_Sensing,birari2025dualdielectric,alexandropoulos2020hardware,R-RIS} enable the efficient estimation of all channels involving a $k$-th RIS (or BD-RIS) locally at its controller. By augmenting this CSI estimation availability with the estimations of all direct BS-UE channels via dedicated control links of substantially lower load than in the previous centralized approach, its $k$-th considered BD-RIS controller can decide, at each time instance $t$, its local configuration $\boldsymbol{\Phi}_k(t)$ as well as a candidate BS precoding matrix $\mathbf{V}_k(t)\triangleq[\mathbf{v}_{k,1}(t),\ldots,\mathbf{v}_{k,N_{\rm ue}}(t)]$ with $\mathbf{v}_{k,n}(t)\in\mathcal{V}$ $\forall k,n$, as follows. 

Let $g_k(\cdot)$ denote the mapping function to be designed at each $k$-th BD-RIS controller with the goal to map the available instantaneous CSI at time $t$ to $\boldsymbol{\Phi}_k(t)$ and $\mathbf{V}_k(t)$, i.e.:
\begin{equation}\label{eq:g_k}
       \{\boldsymbol{\Phi}_k(t),\mathbf{V}_k(t)\} =g_k\left(\mathbf{H}_{\rm D}(t),\mathbf{H}_{1,k}(t),\mathbf{H}_{{\rm I},2,k}(t)\right),
\end{equation}
where $\mathbf{H}_{{\rm I},2,k}(t)\triangleq[\mathbf{h}_{2,1,k}(t),\ldots,\mathbf{h}_{2,N_{\rm ue},k}(t)]\in \mathbb{C}^{N_{\rm ris} \times N_{\rm ue}}$. Let also $g_{K+1}(\cdot)$ represent a fusion function implemented at the BS side with the objective to combine all latter distributedly calculated candidate $\mathbf{V}_k(t)$'s and collected via the control links~\cite{10600711} to decide the final precoding matrix $\mathbf{V}(t)$; in mathematical terms:
\begin{equation}\label{eq:g_k1}
    \mathbf{V}(t)=g_{K+1}\left(\left\{\mathbf{V}_k(t)\right\}_{k=1}^K\right).
\end{equation}
It is noted that, since the precoding codebook $\mathcal{V}$ is available at the BS and can be efficiently stored at all BD-RIS controllers upon their initial installation, the $\mathbf{V}_k(t)$'s distributedly calculated at each time instant $t$ can be made available at the BS via a few bits control signaling mechanism concerning only their indices within the codebook. 

Similar to the previously presented centralized approach, at every time instance $t$, \eqref{eq:g_k} and \eqref{eq:g_k1} can be invoked at each $k$-th BD-RIS controller to design the pair $\{\boldsymbol{\Phi}_k(t),\mathbf{V}_k(t)\}$ upon observing the current relevant CSI and at the BS to design $\mathbf{V}(t)$ upon collecting all distributedly derived $\mathbf{V}_k(t)$'s, respectively. For this Hybrid Distributed and Fusion (HDF) purpose, a sum-SNR maximization objective over a finite time horizon $T$, similar to $\mathcal{OP}_2$, can be adopted which is expressed mathematically as follows:
\begin{align*}
  \mathcal{OP}_3: \max_{\forall g_k(\cdot) \in \mathcal{G}_k} &  {\rm E}\left[ 
  \sum_{t=1}^{T} R\left(\left\{g_k\left(\mathbf{H}_{\rm D}(t),\mathbf{H}_{1,k}(t),\mathbf{H}_{{\rm I},2,k}(t)\right)\right\}_{k=1}^K ,g_{K+1}\left(\left\{\mathbf{V}_k(t)\right\}_{k=1}^K\right)\right) \right]\,\\
  &\textrm{s.t.} \quad ({\rm C1})\,\,{\rm and } \, ({\rm C3}):\mathbf{v}_{k,n}(t)\in\mathcal{V}  \,\,\forall k,n,
\end{align*}
where each $\mathcal{G}_k$ (with $k=1,\ldots,K+1$) represents the set  of all applicable mappings $g_k(\cdot)$. Each of the first $K$ sets contains any function at each of the $K$ BD-RIS controllers capable of mapping local CSI to the feasible sets of the local BD-RIS configurations and the BS precoding matrices resulting from them. The set $\mathcal{G}_{K+1}$ contains any function at the BS capable of mapping the distributedly derived candidate precoding matrices to the feasible sets of the final BS precoding matrix.

\begin{figure}[!t]
    \centering
    \includegraphics[width=0.95\linewidth]{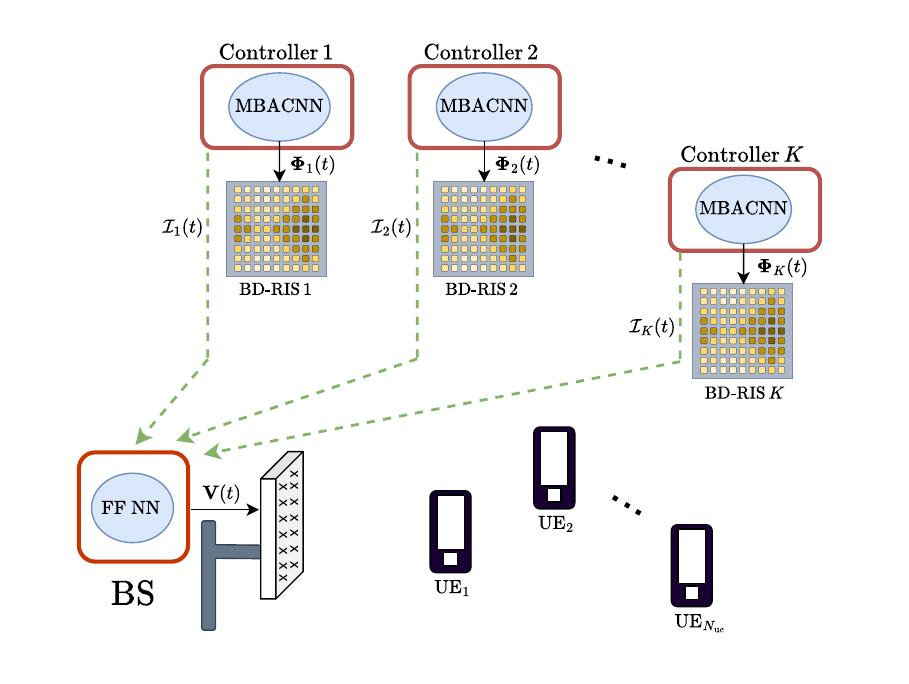}
    \caption{The architecture of the proposed HDF ML framework. Each $k$-th BD-RIS controller designs, via an appropriately designed NN, the configuration $\boldsymbol{\Phi}_k(t)$ of its metasurface panel as well as a candidate index set $\mathcal{I}_k(t)$ from the BS precoder matrix codebook. Those $K$ index sets are then collected, via control links, and fused, through an adequately designed NN, at the BS side to decide the final multi-UE precoding matrix $\boldsymbol{V}(t)$. The details of the proposed Multi-Branch Attention Convolution NN (MBACNN) and the Feed Forward (FF) NN are presented in Section~\ref{sec:HDF}.}
    \label{fig:multi-risSetUP}
\end{figure}

In the following, each local mapping function $g_k(\cdot)$ $\forall k=1,\ldots,K$ is parameterized with an elaborate NN, which is assumed to be implemented at each $k$-th BD-RIS controller (e.g., through a GPU installed to it). Similarly, the BS hosts an elaborate NN parameterizing the mapping function $g_{K+1}(\cdot)$. The optimization of the weights of all those $K+1$ NNs will be conducted offline by simulating large numbers of UE trajectories of length $T$, and then, the expectation over the channels will be approximated by averaging the sums of all trajectories. During the online testing/deployment phase of the $K+1$ optimized distributed NNs, at each channel coherence time instance $t$, the channels involving each $k$-th BD-RIS will be first estimated locally via existing CSI estimation methods, and then, provided to the locally available NN. Each $k$-th of these networks will then quickly decide on the near-optimal pair including the $k$-th BD-RIS configuration and the candidate BS precoder. It is noted that the operation of these $K$ NNs can take place in parallel. Next, all $K$ candidate BS precoding matrices will be gathered and provided at the NN installed at the BS side to quickly decide on the final BS precoding matrix to serve all UEs. This HDF ML framework, whose constituent NNs are detailed in the following Section~\ref{sec:HDF}, is schematically illustrated in Fig.~\ref{fig:multi-risSetUP}.

\subsection{Proposed Hybrid Distributed and Fusion Machine Learning Design}\label{sec:HDF}
This section presents the core components of the HDF ML framework solving $\mathcal{OP}_3$, which comprises $K+1$ NNs as illustrated in Fig.~\ref{fig:multi-risSetUP} and previously described in Section~\ref{sec:distributed}. Each of the $K$ NNs implemented at each $k$-th BD-RIS controller is a Multi-Branch Attention Convolution NN (MBACNN)~\cite{mbacnn} responsible for extracting at each time instance $t$ the $k$-th BD-RIS configuration $\boldsymbol{\Phi}_k(t)$ and the candidate BS precoder $\mathbf{V}_k(t)$. In the sequel, the architecture of the proposed MBACNN model are presented, starting with the motivation behind its core components. Then, the Feed Forward (FF) NN implemented at the BS to decide on the final BS precoding matrix $\mathbf{V}(t)$ is discussed. Finally, the optimization of the parameters of all $K+1$ NNs, which relies on the Cooperative Synapse NeuroEvolution (CoSyNE) algorithm \cite{cosyne} and a parameter sharing mechanism, is introduced.

In order to handle operations with complex numbers within the proposed distributed ML framework, the estimates of the wireless channel gains are transformed before being passed into each $k$-th ($k=1,\ldots,K$) MBACNN model as follows:
\begin{align}
 \tilde{\mathbf{H}}_{\rm D}(t)& \triangleq [\Re\{\mathbf{H}_{\rm D}(t)\}\,\Im\{\mathbf{H}_{\rm D}(t)\}] \in \mathbb{R}^{2N_{\rm tx}\times N_{\rm ue}},\label{eq:input-h}\\
    \tilde{\mathbf{H}}_{1,k}(t)&\triangleq\begin{bmatrix}
        \Re\{\mathbf{H}_{1,k}(t)\} \, \Im\{\mathbf{H}_{1,k}(t)\}
    \end{bmatrix} \in \mathbb{R}^{2N_{\rm tx} \times N_{\rm ris}},\label{eq:input-H1}\\
    \tilde{\mathbf{H}}_{{\rm I},2,k}(t)&\triangleq [\Re\{ \mathbf{H}_{{\rm I},2,k}(t)\} \, \Im\{ \mathbf{H}_{{\rm I},2,k}(t)\}] \in \mathbb{R}^{2N_{\rm ris} \times N_{\rm ue}}.\label{eq:input-h2}
\end{align}
\begin{figure}[!t]
\centering
    \begin{subfigure}[]{\textwidth}
        \scalebox{0.5}{\includegraphics{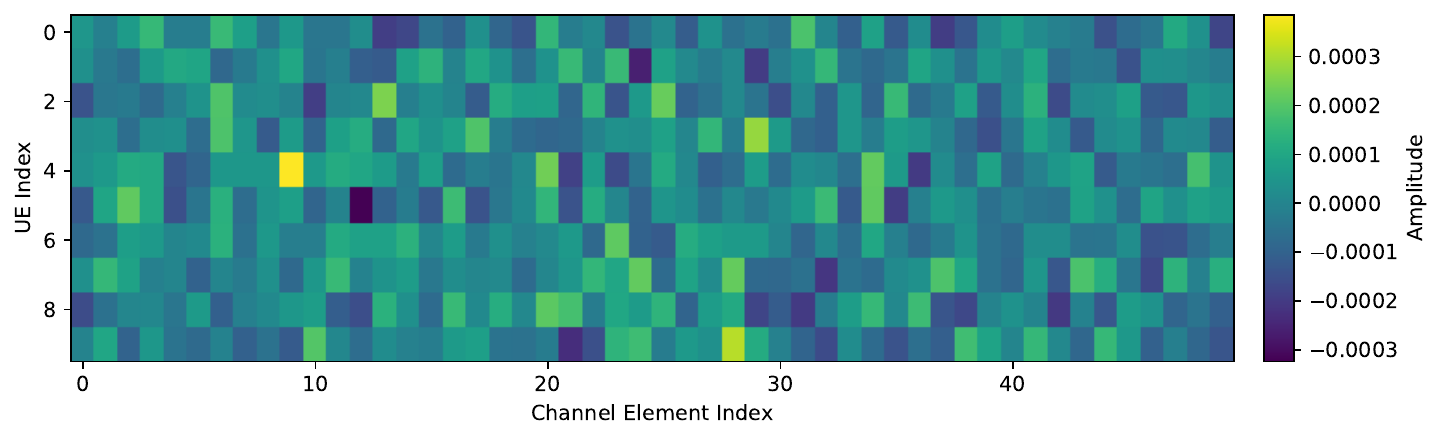}}
    \caption{Real part of a $\mathbf{H}_{1,k}(t)$ realization.}
    \end{subfigure}\\
    \begin{subfigure}[]{\textwidth}
        \scalebox{0.5}{\includegraphics{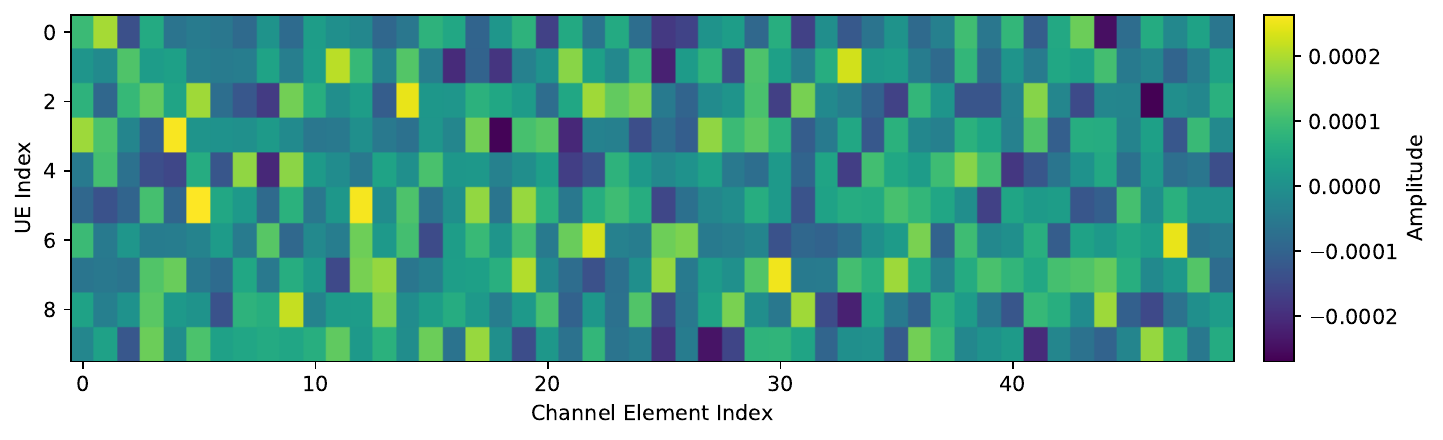}}
    \caption{Imaginary part of a $\mathbf{H}_{1,k}(t)$ realization.}
    \end{subfigure}
    \caption{The amplitudes of the real and imaginary parts of a $10\times50$ channel matrix $\mathbf{H}_{1,k}(t)$ constituting a realization of the Ricean distribution \cite[eq. (5)]{Alexandropoulos2021_PervasiveML} with $7$ dB $\kappa$-factor. It can be observed that adjacent matrix elements have similar values (spatial correlation), a fact that motivates the investigation of attention mechanisms as a means to extract important channel features.}
    \label{fig:riceanSeqDemo}
\end{figure}
\subsubsection{Motivation}
Inspired by the potential of metasurfaces to enable SWEs, and their increased optimization potential especially in low angular spread scenarios~\cite{Moustakas_Cap2023,moustakas2024mimo,Alexandropoulos2021_PervasiveML}, Fig.~\ref{fig:riceanSeqDemo} illustrates the amplitudes of the real and imaginary parts of a realization of a $10\times50$ channel matrix $\mathbf{H}_{1,k}(t)$ (i.e., between a $10$-antenna BS and an BD-RIS with $50$ elements) drawn from the Ricean channel model in \cite[eq. (5)]{Alexandropoulos2021_PervasiveML} with a moderate $\kappa$-factor at $7$ dB. It can be observed that adjacent columns of this matrix have similar values due to the induced spatial correlation~\cite{CorGA2} arising from the steering vector expressions. In fact, each column strongly depends on neighboring columns, implying that the columns of the channel matrix form patterns. This fact motivates the consideration of sequence modeling tools~\cite{attentionSurvey,AttentionAllYouNeed,nueralMachineTranslation}  to extract important information about wireless channels.

\subsubsection{MBACNN Architecture at Each $k$-th BD-RIS Controller for $\boldsymbol{\Phi}_k(t)$}\label{sec:MBACNN_architecture}
\begin{figure*}[!t]
    \centering
    \scalebox{0.5}{\includegraphics{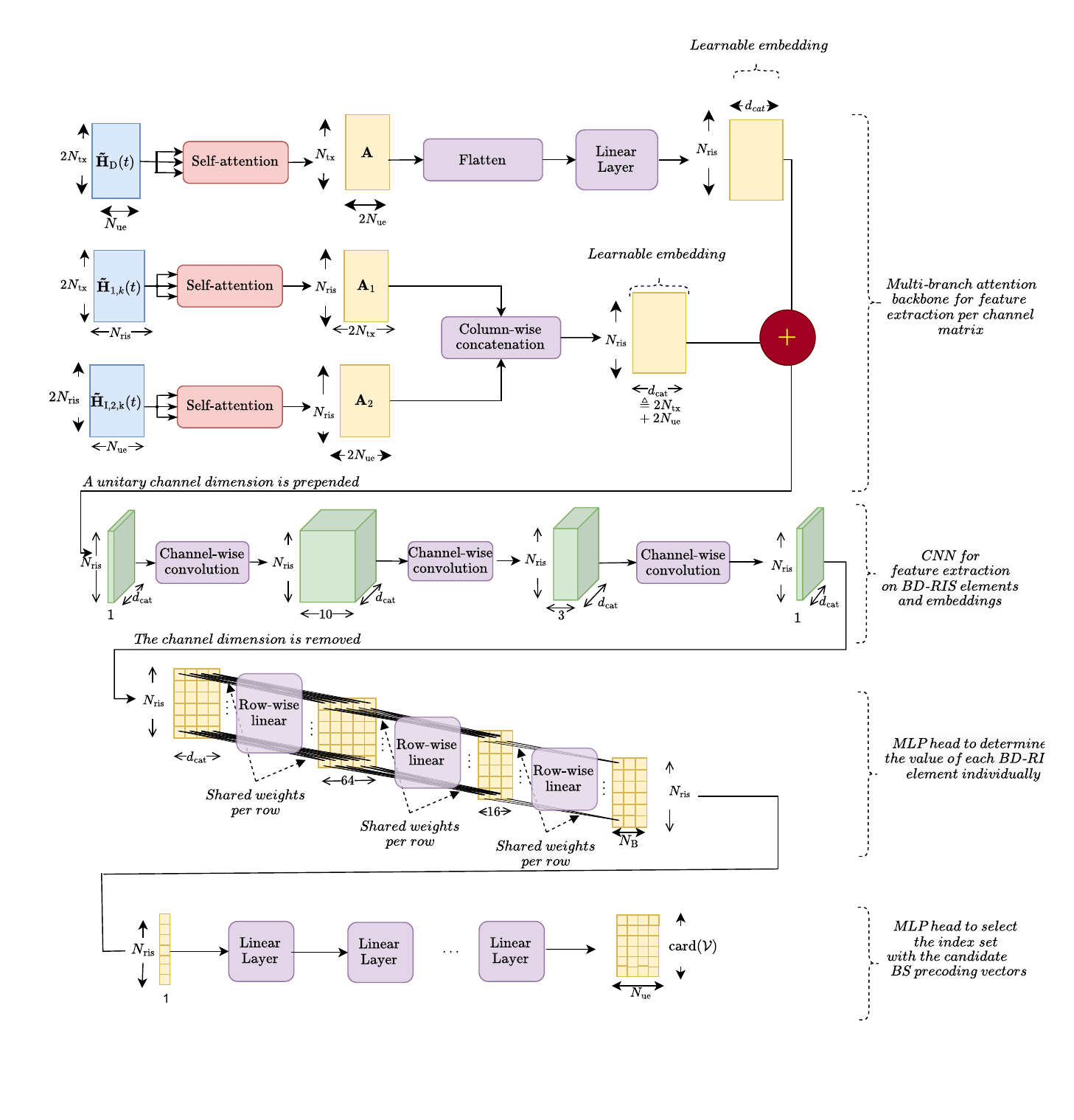}}
    \caption{The proposed MBACNN architecture with its CSI inputs $\tilde{\mathbf{H}}_{\rm D}(t)$, $\tilde{\mathbf{H}}_{1,k}(t)$, and $\tilde{\mathbf{H}}_{\rm I,2,k}(t)$ at each time instance~$t$, which is hosted at each $k$-th BD-RIS controller. This architecture comprises a multi-branch attention module, followed by a convolutional NN module, a Multi-Layered Perceptron (MLP) module for the selection of the BD-RIS configuration matrix $\boldsymbol{\Phi}_k(t)$, and an additional MLP module for selecting the set of indices $\mathcal{I}_k(t)$ indicating the candidate BS precoding matrix $\mathbf{V}_k(t)$.}
    \label{fig:dACNN}
\end{figure*}
The proposed MBACNN architecture illustrated in Fig.~\ref{fig:dACNN} and intended for installation at each $k$-th BD-RIS controller consists of the following modules.

\textbf{Multi-Branch Attention Module:}
The first module receives as inputs the CSI estimates $\tilde{\mathbf{H}}_{\rm D}(t)$, $\tilde{\mathbf{H}}_{1,k}(t)$, and $\tilde{\mathbf{H}}_{1,2,k}(t)$ at each time instant $t$, and is tasked for feature extraction. This module is based on self-attention~\cite{attentionSurvey}, and precisely, the ``Scaled-Dot-Product Attention'' operation of \cite[eq. (1)]{AttentionAllYouNeed}. By viewing each channel matrix as a sequence of (row) vectors, this module seeks patterns among the channel coefficients related to the BD-RIS unit elements. Each sequence comprises unit elements, with each of them represented as a token vector containing the coefficients at the BS and the coefficients at the BS and all UE antennas ($N_{\rm tx}+N_{\rm ue}$ in total) that serve as an embedding. Self-attention architectures are efficient in extracting sequence-related information~\cite{nueralMachineTranslation}, in effect, unveiling the channel conditions that induced the correlations between the channel coefficients~\cite{Nakagami_corr,CorGA1} at adjacent BD-RIS elements.  

As shown in the first module in Fig.~\ref{fig:dACNN}, each of the channels $\tilde{\mathbf{H}}_{\rm D}(t)$, $\tilde{\mathbf{H}}_{1,k}(t)$, and $\tilde{\mathbf{H}}_{1,2,k}(t)$ is passed on its separate self-attention layer in order to capture important features regarding the correlations of the channel matrix elements. Their learnable parameters are intended to weigh the correlations so that the NN identifies which correlations are useful for the active/passive beamforming problem at hand. Consider for example the channel matrix $\tilde{\mathbf{H}}_{1,k}(t)$, for which the time variable notation is removed for simplicity in the following expressions. The first term in implementing neural self-attention for $\tilde{\mathbf{H}}_{1,k}$ is to compute the intermediate attention scores as follows:
\begin{equation}\label{eq:attention}
    \mathbf{S}_1 \triangleq {\rm softmax}\left( \frac{\mathbf{Q}_1 \mathbf{K}_1^{\rm T}}{\sqrt{2N_{\rm tx}}}  \right) \in \mathbb{R}^{{N_{\rm ris} \times N_{\rm ris}}},
\end{equation}
where ${\rm softmax}(\cdot)$ is taken over the complete array to convert all values in the range $[0,1]$, and matrices $\mathbf{Q}_1$ and $\mathbf{K}_1$ are defined as:
\begin{align}\label{eq:attention-Q-K}
    \mathbf{Q}_1 &\triangleq \mathbf{W}^q_1 \tilde{\mathbf{H}}_{1,k} , \quad 
    \mathbf{K}_1 \triangleq \mathbf{W}^k_1 \tilde{\mathbf{H}}_{1,k}  \in \mathbb{R}^{N_{\rm ris} \times 2N_{\rm tx}},
\end{align}
with $\mathbf{W}^q_1$ and $\mathbf{W}^k_1$ are real-valued learnable weight matrices of dimension $N_{\rm ris} \times N_{\rm ris}$. Note that the matrix multiplication inside \eqref{eq:attention} may be interpreted as row-wise dot products that measure the row similarity for all pairs of rows in $\tilde{\mathbf{H}}_{1,k}$. The weight matrices perform task-specific linear transformations on these matrix row pairs. The final, attended output of each attention layer is given as follows:
\begin{equation}\label{eq:attention-output}
    \mathbf{A}_1 \triangleq \mathbf{S}_1 \mathbf{P}_1 \in \mathbb{R}^{ N_{\rm ris} \times 2 N_{\rm tx}},
\end{equation}
where the following matrix definition has been used:
\begin{equation}\label{eq:attention-V}
   \mathbf{P}_1 \triangleq \mathbf{W}^v_1 \tilde{\mathbf{H}}_{1,k}  \in \mathbb{R}^{N_{\rm ris} \times 2 N_{\rm tx}}
\end{equation}
with $\mathbf{W}^v_1 \in \mathbb{R}^{N_{\rm ris}\times N_{\rm ris}}$ being trainable and having the role to further measure the relevance between the attention scores and the task-specific learnable information embedded in its weights. 
\begin{figure}[t]
    \centering
    \includegraphics[width=0.9\linewidth]{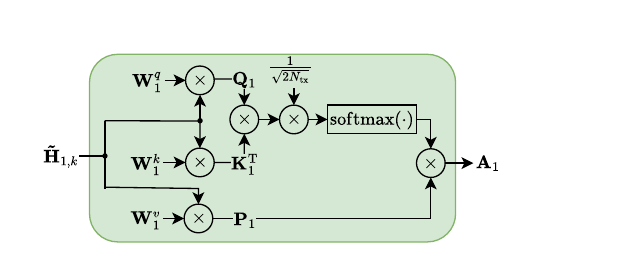}
    \caption{ Block diagram of the scaled-dot-product self-attention layer for feature extraction on the channel matrix $\mathbf{\tilde{H}}_{1,k}$.}
    \label{fig:H-self-attention-block-diagram}
\end{figure}

The overall structure of the proposed self-attention module is given in Fig.~\ref{fig:H-self-attention-block-diagram}. 
Following similar reasoning, the attended outputs $\mathbf{A}_2 \in \mathbb{R}^{N_{\rm ris} \times 2 N_{\rm ue}}$ and $\mathbf{A} \in \mathbb{R}^{N_{\rm tx} \times 2 N_{\rm ue}}$, resulting after respectively passing $\tilde{\mathbf{H}}_{\rm D}$ and $\tilde{\mathbf{H}}_{\rm I,2,k}$ through appropriate self-attention branches, are constructed. The matrices $\mathbf{A}_1$ and $\mathbf{A}_2$ are concatenated along their column dimension in the matrix $\mathbf{A}_{\rm C} \triangleq {\rm colcat}(\mathbf{A}_1, \mathbf{A}_2) \in \mathbb{R}^{N_{\rm ris} \times d_{\rm cat}}$ with $d_{\rm cat} \triangleq 2N_{\rm tx} + 2 N_{\rm ue}$.
In order to be able to merge all the matrices before passing them to the next module in the proposed MBACNN architecture, an additional FF layer with the appropriate number of hidden units is utilized to transform $\mathbf{A}$ (first flattened to an $2N_{\rm tx} \times 1$ vector) into the $\mathbf{A}_0 \in \mathbb{R}^{d_{\rm cat} N_{\rm ris} \times 1}$ matrix. This matrix is then reshaped to the dimension $N_{\rm ris} \times d_{\rm cat}$, and the final attended matrix is obtained as follows:
\begin{equation}\label{eq:A_T}
\mathbf{A}_{\rm T} \triangleq \mathbf{A}_{\rm C}+\mathbf{A}_0,    
\end{equation}
where layer normalization needs to be performed at each of these two terms. It is finally noted that, in many RIS-focused investigations (e.g.,~\cite{Stylianop1Bit,MIMOris1bit,Alexandropoulos2021_PervasiveML}), the direct channel $\mathbf{H}_{\rm D}(t)$ is considered as blocked or highly attenuated. This case can be easily treated by the proposed MBACNN architecture: the architecture in Fig.~\ref{fig:dACNN} simplifies by removing the attention head that receives as input the $\tilde{\mathbf{H}}_{\rm D}(t)$ matrix.

\textbf{Convolutional NN Module:} The next module for feature extraction is based on cascaded convolutional layers. To this end, the attended matrix $\mathbf{A}_{\rm T}$ formulated in \eqref{eq:A_T} is treated as an image-like tensor of dimension $1 \times N_{\rm ris} \times d_{\rm cat}$, where a single image channel is implied during the first dimension. Three convolutional layers are applied to the input tensor with the hyper-parameters configured so that the last two output dimensions of each layer remain unchanged. Instead, the channel dimension (i.e., number of kernels) differs among the layers and is used to allow the NN to store arbitrary information. The output of the final convolutional layer again consists of a single channel, which is then discarded to construct a matrix of the same dimension as $\mathbf{A}_{\rm T}$. 

The convolutional layers of this module are intended to extract spatial patterns in the ``image-like'' stacked attention matrix, as well as to decrease the dimensionality of the input. In fact, notwithstanding rich scattering conditions, channel matrices exhibit high spatial locality in terms of phase values and possibly amplitude, therefore, the use of convolutional kernels is expected to be highly effective in detecting them.

\textbf{Multi-Layered Perceptron Module for the BD-RIS Configuration $\boldsymbol{\Phi}_k(t)$:} Having used the attention and convolution layers to extract geometric information features, FF layers are adopted to map said features to BD-RIS profiles. A BD-RIS with $N_{\rm ris}$ unit elements of each with $b$ bits resolution in the response state and $N_{\rm B}$ superdiagonals and subdiagonals necessitates an output branch comprising a sequence of Rectified Linear Unit (ReLU) activated layers, configured such that the final vector output dimension is $(2N_{\rm B}+1)N_{\rm ris}$. The first $N_{\rm ris}$ entries of this vector correspond to the main diagonal of $\boldsymbol{\Phi}_k(t)$ (responses of meta-atoms, i.e., phase profile) and the rest to its $N_{\rm B}$ superdiagonals and subdiagonals (configurations of the ON/OFF element connection switches). For example, for $b=1$ and the binary states $\theta_1$ and $\theta_2$, a $\tanh(\cdot)$ and a $\text{sign}(\cdot)$ function are then applied to transform the $N_{\rm ris}$ entries of this vector, corresponding to the meta-atom responses, into an equal size vector having either $\theta_1$- or $\theta_2$-valued elements. The same applies to the remaining $2N_{\rm B}N_{\rm ris}$ entries of that vector that need to be mapped to binary values corresponding to the ON and OFF switch states (e.g., ``1'' for ON and ``0'' for OFF). Note that, for BD-RISs with $b\geq2$, the $\tanh(\cdot)$ and $\text{sign}(\cdot)$ operations for only the latter $N_{\rm ris}$ entries need to be replaced by the ${\rm softmax}(\cdot)$ function to decide the final allowable element response states.

\textbf{Multi-Layered Perceptron Module for the Candidate BS Precoder $\mathbf{V}_k(t)$:} The last module of the MBACNN architecture is responsible for deciding $\mathbf{v}_{k,n}(t)$'s (i.e., the BS precoding vector for each $n$-th UE) solving $\mathcal{OP}_3$, while each belonging in the discrete codebook $\mathcal{V}$ (i.e., the (C3) constraint in $\mathcal{OP}_3$). To this end, as shown in Fig.~\ref{fig:dACNN}, $N_{\rm ue}$ parallel stacks of linear layers (or a multi-head structure) are used, with each corresponding to a distinct UE and intended to map the extracted features to an output vector of size ${\rm card}(\mathcal{V})\times 1$. This output of each stack is then passed through a ${\rm softmax}(\cdot)$ activation function to generate a discrete distribution over all possible precoding vectors for that specific UE. Then, a precoding vector preference $\mathbf{v}_{k,n}(t)$ for each UE is sampled from these distributions. Since $\mathcal{V}$ is discrete and finite, this selection corresponds to a set of indices $\mathcal{I}_k(t) \triangleq \{i_{k,1}(t), \dots, i_{k,N_{\rm ue}}(t)\}$, with each $i_{k,n}(t)$ indicating an integer between the values of $1$ and ${\rm card}(\mathcal{V})$. 

\subsubsection{Feed Forward NN at the BS for $\mathbf{V}(t)$}
The preference indices $\mathcal{I}_k(t)$'s for the BS precoder vectors per UE, computed via the proposed MBACNN architecture at each $k$-th BD-RIS controller, are transferred through the $K$ control links to the BS~\cite{10600711,KunzBaskaranAlexandropoulos2025}. At this node, these candidate precoders via their indices are aggregated into a global preference matrix and then passed through a small FF NN. This network utilizes a ${\rm softmax}(\cdot)$ activated output layer to decide the final codebook indices determining the final precoding matrix $\mathbf{V}(t)$, where each column $[\mathbf{V}(t)]_{:,n}$ represents the precoding vector to be finally applied when transmitting actual data to each $n$-th UE in the downlink direction.

\subsubsection{Evolutionary Optimization of the NN Parameters}\label{sec:NE_training}
All $K$ BD-RIS controllers are assumed to operate with an identical set of MBACNN weights, which is denoted by the matrix $\mathbf{W}_{\rm ris}$. In addition, the stacked learnable weights of the FF NN at the BS are represented by the matrix $\mathbf{W}_{\rm bs}$. Let also $g_{\mathbf{W}_{\rm ris}}(\cdot)$ and $g_{\mathbf{W}_{\rm bs}}(\cdot)$ represent the control policies governed by the NNs with the parameter matrices $\mathbf{W}_{\rm ris}$ and $\mathbf{W}_{\rm bs}$, respectively. It is noted that, by virtue of the previously presented NN output layers and the employment of the $\tanh(\cdot)$ and $\rm{softmax}(\cdot)$ activation functions, the generated BD-RIS configurations will meet constraint (C1). Similarly, the selected BS precoding vectors are guaranteed to reside within the codebook $\mathcal{V}$ (i.e., constraints (C2) and (C3)).  

A primary challenge associated with $\mathcal{OP}_3$ is that its objective (i.e., the ergodic sum-rate performance) lacks a closed-form expression. This fact necessitates simulating an infinite number of channel realizations for its exact evaluation. To address this intractability, rather than optimizing the mappings on the exact objective, the NN parameters governing the mappings are optimized using a sample average approximation over $N_{\mathrm{EP}}$ trajectories. Let $\mathbf{W} \triangleq [\mathbf{W}_{\mathrm{ris}}\, \mathbf{W}_{\mathrm{bs}}]$ denote the concatenated set of learnable NN parameters and $R_p(t;\mathbf{W})$ represent, at time instant $t$, the achievable sum rate of the $p$-th trajectory set for all $N_{\rm ue}$ UEs, given that the BD-RIS configurations and the BS precoding matrix are determined by the policies $g_{\mathbf{W}_{\rm ris}}(\cdot)$ and $g_{\mathbf{W}_{\rm bs}}(\cdot)$, respectively. To design $\mathbf{W}$, the following optimization problem is formulated:
\begin{equation}
    \mathcal{OP}_4 : \max_{\mathbf{W}} 
    \sum_{p=1}^{N_{\rm EP}} \sum_{t=1}^T R_p(t;\mathbf{W}).
\end{equation}

To solve the latter problem jointly for all the NN parameters $\mathbf{W}$, neuroevolution~\cite{stamatelisPowerControl,salimans2017evolution,introtoEA,eaht}, and specifically the CoSyNE algorithm~\cite{cosyne}, is adopted. In particular, the evaluation of each individual candidate solution is conducted by splitting it into two NNs: one common NN to be installed at all BD-RIS controllers and the other NN to be hosted at  the BS. For each $p$-th  episode spanning a time horizon $T$, channel realizations are sampled and distributed, at each time step $t$, to all BS-RIS controllers. Subsequently, each $k$-th controller's MBACNN utilizes the parameters $\mathbf{W}_{\rm ris}$ to determine the BD-RIS configuration matrix $\boldsymbol{\Phi}_k(t)$ and the candidate BS precoding indices $\mathcal{I}_k(t)$, transmitting the latter to the BS via the respective control channel. Then, the BS uses the parameters $\mathbf{W}_{\rm bs}$ to design the final precoding matrix $\mathbf{V}(t)$. The resulting instantaneous achievable sum-rate performance is computed and accumulated to derive the sampled average sum-rate metric. In particular, after evaluating $\mathcal{OP}_4$'s objective function for all individuals, the population is ranked by performance~\cite{cosyne}. Following the  CoSyNE algorithm, the weights of the elite candidates then serve as parents to generate the offspring for the next generation using crossover and mutation. Note that the optimization of both NN parameter matrices $\mathbf{W}_{\rm ris}$ and $\mathbf{W}_{\rm bs}$ happens simultaneously by the same evolutionary optimization algorithm.


\subsection{Numerical Results and Discussion}
In this section, performance evaluation results for the HDF ML approach described in Section~\ref{sec:HDF} are presented. Various RIS-enabled broadcast SWEs have been simulated:
\begin{itemize}
    \item A single-RIS ($K=1$) SWE where the RIS was placed in LoS with the BS, the channel between the single ($N_{\rm ue}=1$) UE and the RIS was modeled as a Ricean fading one with factor $\kappa=10$~dB, and the direct BS-UE channel was assumed totally blocked due to the presence of obstacles. The BS was positioned at the Cartesian coordinates $(0, 0, 2.0)$~m, while the UE was placed at the position $(8,10,1.5)$~m. Both a conventional diagonal RIS (i.e., $N_{\mathrm{B}}=0$) and BD-RISs (i.e., $N_{\mathrm{B}}=\{1,2\}$) have been considered with any of them positioned at the point $(0,3,2.0)$~m.
    \item A single-RIS ($K=1$) SWE with multiple ($N_{\rm ue}>1$) UEs whose position was sampled, at each time instance $t$, from the distribution $\mathcal{N}([9.3, 14.9, 2.1]^{\rm T},\mathbf{I}_{3})$. Similar fading conditions with the previous SWE were considered, with the only difference that, in this case, each direct BS-UE channel was modeled as Ricean faded with factor $\kappa=10$~dB exhibiting also an attenuation of $10$~dB.
    \item An SWE with $K=4$ conventional diagonal RISs ($N_{\rm B}=0$) positioned at the points $(3, 3, 2)$~m, $(6,6,-2)$~m, $(3, 3, -2)$~m, and $(6,6,2)$~m and serving a single ($N_{\rm ue}=1$) UE in the downlink direction, which is placed at the position $(8,10,1.5)$~m as in the first SWE. The fading conditions were the same with the previous SWE.
\end{itemize}
In all considered RIS-empowered downlink
MISO systems, the BS possessed $N_{\rm tx}=16$ antenna elements forming a uniform linear array. In addition, each RIS was modeled as a uniform planar array of $N_{\rm ris}=20\times20$ unit elements, and the noise level at each single-antenna UE was set to $-50$ dBm. 

For the implementation of the CoSyNE algorithm~\cite{cosyne} solving $\mathcal{OP}_4$, the mutation probability and variance were set to $0.3$ and $0.2$, respectively. 
A population of $L_{\rm pop}=100$ candidate solutions was evolved for $N_{\rm gen}=25$ generations. For each individual, $\mathcal{OP}_4$'s objective was estimated by simulating $N_{\rm EP}=100$ episodes of horizon $T=50$ (i.e., $5000$ channel realizations in total). The parameters of the best-performing individual of the final generation were finally split into the NN parameters $\mathbf{W}_{\rm ris}$ for each of the RIS controllers and the NN parameters $\mathbf{W}_{\rm bs}$ at the BS, as previously explained. For the overall training process, $20$ runs of different initialization seeds were averaged.

For the average rate and sum-rate performance results that follow, $5000$ MC samples were used as inputs to the proposed NE-trained HDF ML approach during its deployment phase. For comparison purposes, for the simulated SWEs with a single conventional diagonal RIS, the following classic benchmarks have been also implemented:
\begin{itemize}
    \item \textbf{Advantage Actor Critic (A2C):} This is a popular policy-based deep reinforcement learning algorithm that can be used for the RIS phase configuration design~\cite{asynchronousForDRL}.
    \item \textbf{NE with large dense FF NNs (NE-FF):} To investigate whether the performance improvement is attributed to the NE-based training algorithm presented in Section~\ref{sec:NE_training} or to the MBACNN architecture detailed in Section~\ref{sec:MBACNN_architecture}, a simple FF NN was trained with the same evolutionary algorithm.
    \item \textbf{Lightweight Genetic Algorithm (LGA):} PyGAD~\cite{pygad} with $L_{\rm pop}^{\rm LGA}=15$ individuals representing candidate RIS phase configurations for $N_{\rm gen}^{\rm LGA}=5$ generations was deployed, at each time instance $t$, for each possible BS precoding matrix to find the best pair for the RIS-enabled SWE design.
    \item \textbf{Block Exhaustive Search (BES):} This method divides the vector in the main diagonal of the single-RIS phase configuration matrix $\boldsymbol{\Phi}_1(t)$ into $N_{\rm blk}$ blocks, where all elements within each block are forced to have the same optimized state. This is a practical RIS case including grouped elements in terms of common phase configuration~\cite{Alexandropoulos2023_RISDeployment}. At each time instance $t$, a brute-force search is performed over all possible BS precoding and grouped block RIS configurations. It is noted that, due to the exponential computational complexity, this benchmark is only applied in single-RIS scenarios with unit elements of $1$-bit response states. In the respective results, $N_{\rm blk}=10$ was used, yielding $2^{10} \times {\rm card}(\mathcal{V})$ pairs of RIS phase configurations and BS precoders at each time instance $t$, for the sum-rate performance computation.
\end{itemize}

\subsubsection{Results for the Single-RIS ($K=1$) SWE with $N_{\rm ue}=1$ UE}
\begin{figure}[!t]
    \centering
    \includegraphics[width=0.7\linewidth]{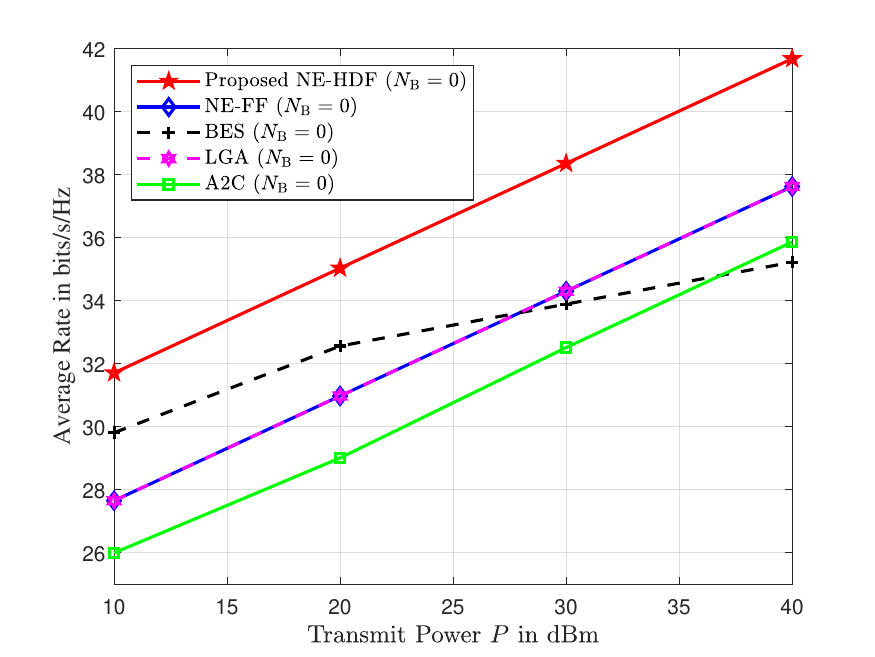}
    \caption{Achievable average rate performance in bits/s/Hz versus the transmit power $P$ in dBm with the presented HDF approach, trained via NE as presented in Section~\ref{sec:NE_training}, and the considered benchmark schemes, considering a single ($K=1$) conventional RIS ($N_{\rm B}=0$) with $N_{\rm ris}=400$ uncoupled unit elements, a BS with $N_{\rm tx}=16$ transmit antennas, and a single ($N_{\rm ue}=1$) UE with noise level of $-50$ dBm. All RIS-aided wireless fading channels were simulated as Ricean distributed with $\kappa=10$~dB, while the direct BS-UE channel was blocked.}
    \label{fig:MBACCN-diagonal}
\end{figure}

The average rate performance results in Fig.~\ref{fig:MBACCN-diagonal} for the conventional diagonal RIS case ($N_{\mathrm{B}}=0$) showcase that the proposed NE-HDF approach outperforms all benchmarks for all considered transmit power values $P$. It can be particularly observed that NE-HDF significantly outperforms its NE-FF variant, indicating that there is a clear benefit in utilizing the proposed MBACNN architecture to extract channel correlation features. It is also demonstrated that A2C performs poorly over the entire $P$-range, while the BES scheme is particularly weak for moderate-to-large $P$ values. The latter behavior can be attributed to the fact that block-based optimization is rather crude, and more refined approaches are required in order to fully profit from large transmit power levels.

\begin{figure}[H]
    \centering
    \includegraphics[width=0.7\linewidth]{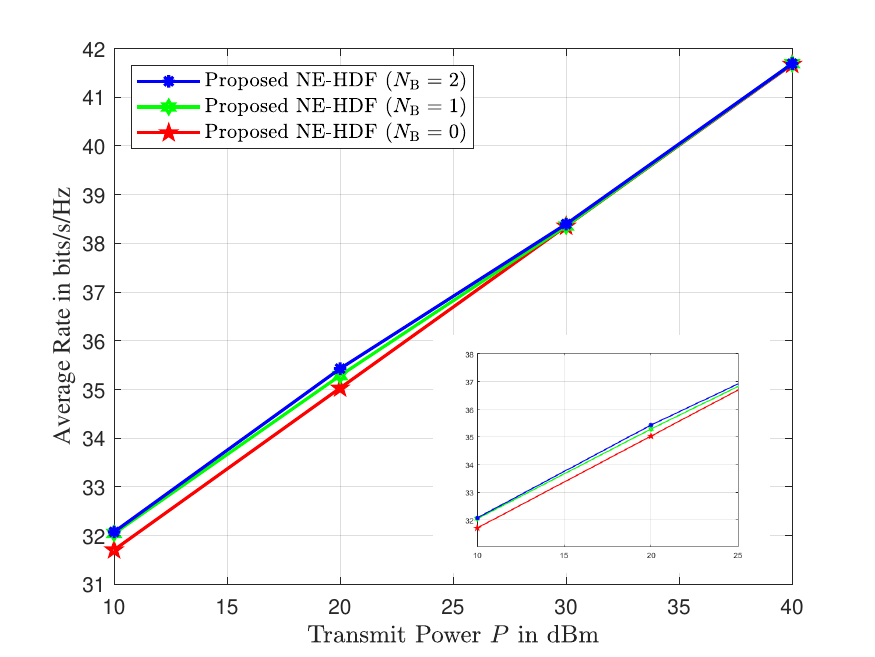}
    \caption{Achievable average rate performance in bits/s/Hz versus the transmit power $P$ in dBm with the presented NE-HDF approach, considering a BS with $N_{\rm tx}=16$ transmit antennas, a single ($N_{\rm ue}=1$) UE with noise level of $-50$ dBm, and different versions of the single-RIS SWE: conventional diagonal RIS ($N_{\rm B}=0$) and BD-RIS ($N_{\rm B}=\{1,2\}$). As in Fig.~\ref{fig:MBACCN-diagonal}, all RIS-aided wireless fading channels were simulated as Ricean distributed with $\kappa=10$~dB, while the direct BS-UE channel was blocked. }
    \label{fig:singleUE-BD}
\end{figure}

Figure~\ref{fig:singleUE-BD} compares NE-HDF's achievable average rate performance in Fig.~\ref{fig:MBACCN-diagonal} (i.e., with the conventional diagonal RIS case ($N_{\mathrm{B}}=0$)) with the cases where a BD-RIS with $N_{\mathrm{B}}=1$ and $N_{\mathrm{B}}=2$ is used. As depicted, the BD-RIS structure provides a modest yet discernible performance improvement when low-to-moderate $P$ values are used. Conversely, in the regime of high transmit power, this performance gap diminishes, suggesting that the conventional diagonal RIS configuration remains a sufficiently robust design choice. This convergence is expected because the logarithmic nature of the achievable rate at high signal-to-noise ration levels naturally suppresses the relative impact of the marginal beamforming gains offered by the increased degrees of freedom provided by the BD-RIS-induced ON/OFF mutual couplings.

\subsubsection{Results for the Single-RIS ($K=1$) SWE with $N_{\rm ue}>1$ UEs}
\begin{figure}[!t]
    \centering
    \includegraphics[width=0.7\linewidth]{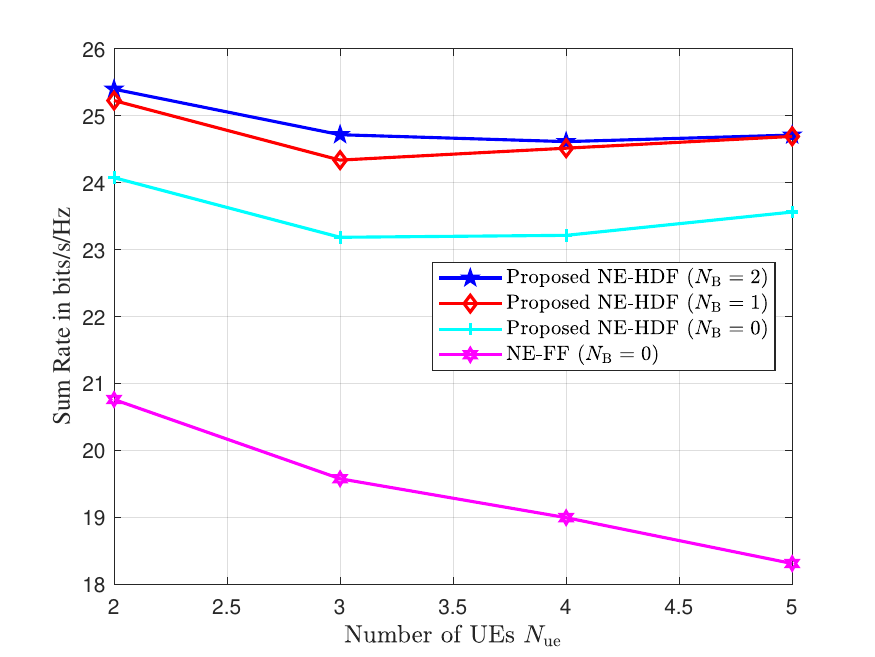}
    \caption{Achievable average sum-rate performance in bits/s/Hz versus the number $N_{\rm ue}$ of UEs with the presented NE-HDF approach, considering the transmit power $P=0.01$~dBm, a BS with $N_{\rm tx}=16$ transmit antennas, noise level at each UE of $-50$ dBm, and either a conventional diagonal RIS ($N_{\rm B}=0$) or a BD-RIS ($N_{\rm B}=\{1,2\}$) with $N_{\rm ris}=400$ unit elements. All RIS-aided wireless fading channels were simulated as Ricean distributed with $\kappa=10$~dB similar to Fig.~\ref{fig:MBACCN-diagonal}, while each direct BS-UE channel was considered as Ricean faded with $\kappa=10$~dB exhibiting an attenuation of $10$~dB.} 
    \label{fig:singleRISmultiUE}
\end{figure}
The average sum-rate results in Fig.~\ref{fig:singleRISmultiUE} showcase that the performance with the proposed NE-HDF approach increases when the SWE is equipped with a BD-RIS with larger $N_{\rm B}$ values. In fact, both considered BD-RIS structures yield superior performance than the conventional diagonal RIS. This enhanced performance can be attributed to the unique capability of the BD-RIS structure to handle inter-UE interference more efficiently than the conventional one, thereby unlocking higher spectral efficiency in RIS-enabled broadcast SWEs. It is finally shown that all NE-HDF versions significantly outperforms their NE-FF variant, indicating again that there is a clear benefit in
utilizing the proposed MBACNN architecture to extract channel correlation features, which increases with increasing $N_{\rm ue}$ values.

\subsubsection{Results for the Multi-RIS ($K>1$) with $N_{\rm ue}=1$ UE}
\begin{figure}[!t]
    \centering
    \includegraphics[width=0.7\linewidth]{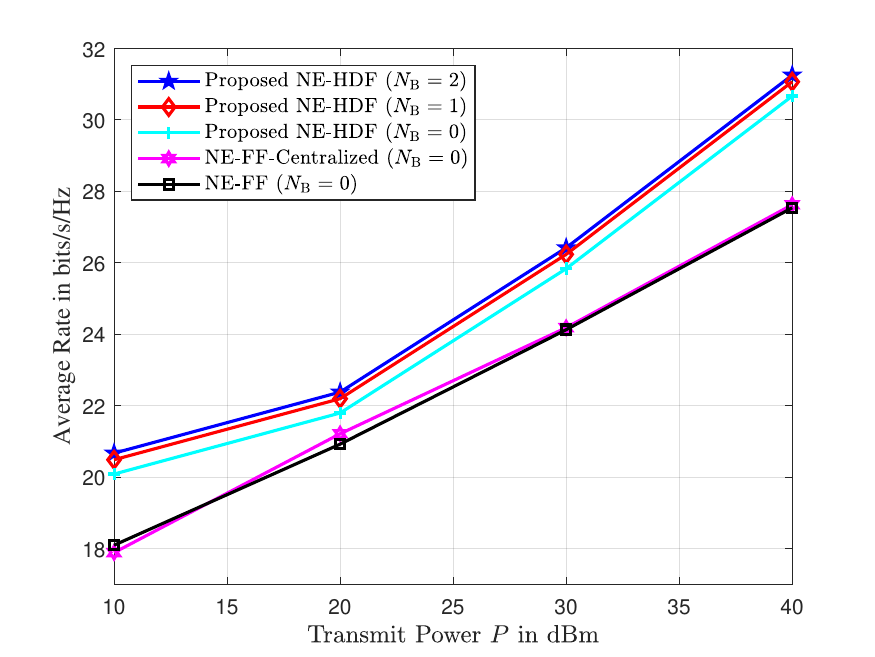}
    \caption{Achievable average rate performance in bits/s/Hz versus the transmit power $P$ in dBm with the presented NE-HDF approach, considering a BS with $N_{\rm tx}=16$ transmit antennas, a single ($N_{\rm ue}=1$) UE with noise level of $-50$ dBm, and $K=4$ conventional diagonal RISs ($N_{\rm B}=0$) with $N_{\rm ris}=400$ uncoupled unit elements. All wireless fading channels including an RIS were simulated as Ricean distributed with $\kappa=10$~dB similar to Fig.~\ref{fig:MBACCN-diagonal}, while the direct BS-UE channel as Ricean fading with $\kappa=10$~dB exhibiting an attenuation of $10$~dB.}
    \label{fig:multi-RIS}
\end{figure}

The proposed NE-HDF method has been also compared against a centralized NE-FF dictating all diagonal RIS phase configurations; this version is termed as ``NE-FF-Centralized'' in Fig.~\ref{fig:multi-RIS}. This NN implementation requires the availability of all SWE's CSI matrices, and when trained, it outputs all $K$ RIS phase configurations as well as the BS precoding vector for the single UE. As is apparent from Fig.~\ref{fig:multi-RIS}, the centralized NE-FF performs poorly because the massive action and observation spaces make it very challenging to represent and optimize the overall system design policy. It can be also observed that, as in the previous figures, the proposed MBACNN in the NE-HDF approach greatly outperforms its distributed FF counterparts. When examining BD-RISs, it is evident that they can achieve a consistent improvement over diagonal ones when controlled by the proposed distributed MBACNN models. In multi-RIS scenarios, the BD-RIS architecture is particularly advantageous as its inter-element coupling unlocks degrees of freedom unavailable in standard diagonal RIS designs. Due to this flexibility, the collaborative beamforming policy learned by the distributed MBACNN agents can exploit the additional mutual couplings to orchestrate a more constructive superposition of signals at the single UE, leading to higher average rate.

\section{Optimization of RIS-Enabled Interference Broadcast SWEs} \label{Sec:IBC_RISs}
In this section, a wideband interference MISO broadcast system operating inside the area of influence of a SWE comprising multiple BD-RISs is optimized for sum-rate maximization. By accounting for the frequency-selective behavior of each RIS unit element, a parallel cooperative scheme that jointly designs the precoding vectors at the multiple multi-antenna BSs as well as the configurations at the BD-RISs (i.e., unit elements' EM responses and their ON/OFF switch interconnections) is presented.

\subsection{System Model} \label{Sec:Sys_Model}
\begin{figure}[!t]
	\centering
	\includegraphics[width=4in]{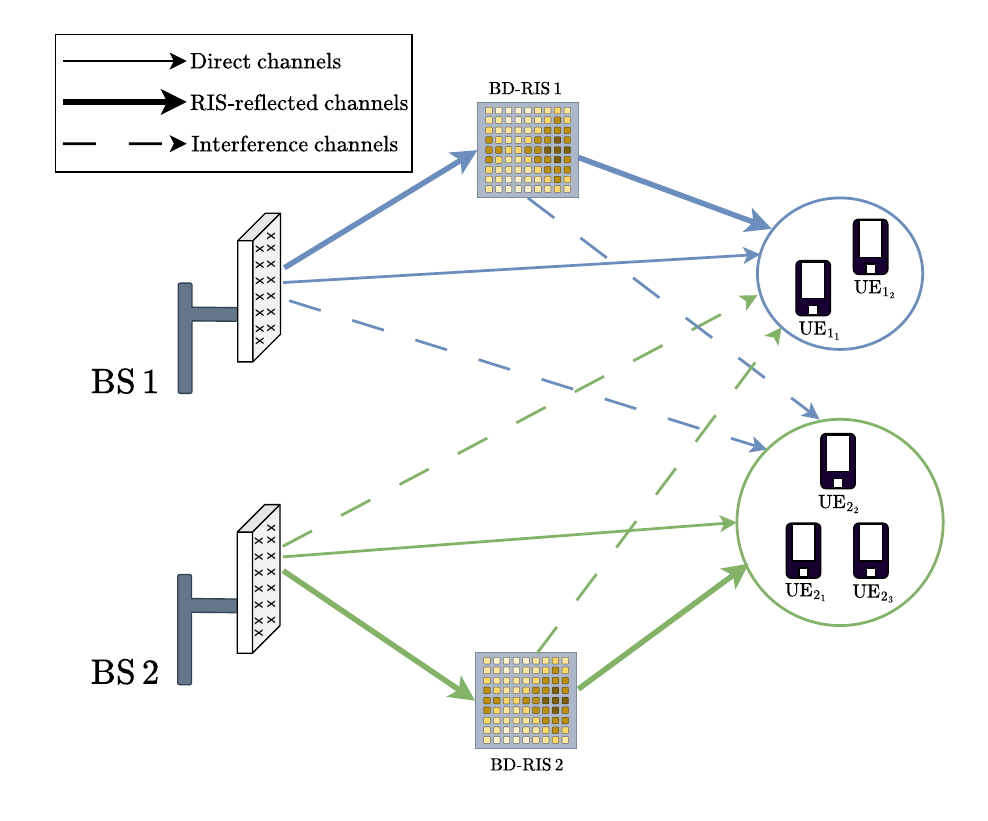}
	\caption{The considered multi-BD-RIS-empowered wireless communication system consisting of $K$ multi-antenna BSs each communicating in the downlink direction with a separate cluster of multiple single-antenna UEs; the case of $K=2$ BSs is illustrated where the first and second cells include $L_1=2$ and $L_2=3$ UEs, respectively. Each $k$-th BD-RIS is asssumed exclusively controlled by its associated $k$-th BS. Propagation paths bouncing over more than one BD-RIS are assumed highly attenuated and are, therefore, neglected.}
	\label{fig:System_Model_IBC}
\end{figure}

An interference broadcast system comprising $K$ multi-antenna BSs each wishing to communicate in the downlink direction with multiple single-antenna UEs through the assistance of $K$ identical BD-RISs is considered, as illustrated in Fig.~\ref{fig:System_Model_IBC}. Each BS equipped with $N_{\rm tx}$ antenna elements is assumed to transmit information to its exclusively associated UEs using Orthogonal Frequency Division Multiplexing (OFDM) in a common set of physical resources, e.g., time and bandwidth. Each $k$-th BD-RIS, comprising $N_{\rm ris}$ unit elements, is assumed to be controlled by its solely owned BS and is placed either closely to it or near to the corresponding cluster of UEs~\cite{Alexandropoulos2023_RISDeployment}. Each BS-UE communicating pair is modeled as the superposition of a direct BS-UE link and a cascaded BS-RIS-UE link realized via the BD-RIS tunable configuration. 
According to the deployed OFDM scheme, the total bandwidth (${\rm BW}$) is equally divided into orthogonal Sub-Carriers (SC) around the carrier frequency $f_c$, with the central frequency of each $n$-th SC ($n=1,\ldots,N_{\rm sub}$) defined as follows:
\begin{equation}
  f_n\triangleq f_c + \frac{\rm BW}{N_{\rm sub}}\left(n - \frac{N_{\rm sub}+1}{2}\right). 
\end{equation}
Let $\vect{w}_{\ell_k,n} \in \mathbb{C}^{N_{\rm tx} \times 1}$ ($k=1,\ldots,K$) represent the precoding vector at each $k$-th BS, which is applied to the unit-power signal $s_{\ell_k,n}$ (i.e., ${\rm E}[|s_{\ell_k,n}|^2] = 1$) before transmission to the $\ell_k$-th UE. By using notation $L_k$ to represent the number of assigned UEs to each $k$-th BS, the corresponding transmit signal $\vect{x}_{k,n}$ can be compactly expressed as $\vect{x}_{k,n} = \sum_{\ell=1}^{L_k} \vect{w}_{\ell_k,n} s_{\ell_k,n}$. Let also $P_k^{\max}$ denote the total transmit power available at each $k$-th BS, hence, the condition $\sum_{\ell=1}^{L_k}\sum_{n=1}^{N_{\rm sub}} \Vert\vect{w}_{\ell_k,n}\Vert^2 \leq P_k^{\max}$ must be satisfied. 

Finally, a quasi-static block fading channel model is assumed for all wireless channels involved, according to which fading changes independently from one block to the next one. The design framework that follows focuses on each particular block where the channels remain approximately constant, assuming also that perfect CSI knowledge is available (see \cite{Jian2022_RIS_Hardware,HRIS_CE_all,RIS_Vahid_low,Swindlehurst_CE} and references therein for CSI estimation frameworks). 

\subsubsection{BD-RIS Structure and Element Response} \label{Sec:Freq_Response}
A BD-RIS hardware architecture incorporating a $N_{\rm ris} \times N_{\rm ris}$ grid of ON/OFF switches, which are intended to interconnect all $N_{\rm ris}$ response-tunable unit elements, is considered. Recall that the ON state at the switch in the $(i,j)$-th grid position ($i,j = 1,\ldots,N_{\rm ris}$) signifies that the signal received by the $i$-th unit element is routed to, and controllably reflected by, the $j$-th element. This switching behavior is captured by a selection matrix $\vect{S}_k \in \{0,1\}^{N_{\rm ris} \times N_{\rm ris}}$ ($k=1,\ldots,K$), which represents the switch configuration of each $k$-th BD-RIS. In particualr, each $\vect{S}_k$ is binary-valued (i.e., $[\vect{S}_k]_{i,j} \in \{0,1\}$) and, by construction, has exactly one non-zero entry in every row and every column, thereby constituting an additional design variable. Note that the conventional diagonal RIS without switches is obtained by setting $\vect{S}_k = \mathbf{I}_{N_{\rm ris}}$.

Each $m$-th unit element ($m=1,\ldots,N_{\rm ris}$) of each $k$-th BD-RIS can be characterized as an equivalent parallel resonant circuit comprising a resistor $R$, a tunable capacitor $C_{mk}$, and two inductors $\mathpzc{L}_1$ and $\mathpzc{L}_2$ \cite{abeywickrama2020intelligent_all}. Then, the response of each $m$-th element of each $k$-th BD-RIS is given by the reflection coefficient which is mathematically expressed in the frequency domain as follows:
\begin{equation} \label{eqn:reflect_coeff}
	\phi_{mk}(f_n,C_{mk}) = \frac{\mathcal{Z}(f_n,C_{mk}) - \mathcal{Z}_0}{\mathcal{Z}(f_n,C_{mk}) + \mathcal{Z}_0},
\end{equation}
where $\mathcal{Z}_0$ is the free space impedance, while $\mathcal{Z}(f_n,C_{mk})$ denotes the characteristic impedance of the equivalent circuit which is given, for $\kappa\triangleq2\pi$, by
\begin{equation} \label{eqn:characteristic_impedance}
	\mathcal{Z}(f_n,C_{mk}) = \frac{\jmath \kappa f_n \mathpzc{L}_1\left(\jmath \kappa f_n \mathpzc{L}_2 + R + \frac{1}{\jmath \kappa f_n C_{mk}}\right)}{\jmath \kappa f_n \left(\mathpzc{L}_1+\mathpzc{L}_2\right) + R + \frac{1}{\jmath \kappa f_n C_{mk}}}.
\end{equation}
Instead of constructing fitting functions to simplify the manipulations of the latter highly non-linear function with respect to the tunable parameters $C_{mk}$, it can be observed that \eqref{eqn:reflect_coeff} can be equivalently transformed into the following more tractable form~\cite{Katsanos_distributed_all}:
\begin{equation} \label{eqn:RIS_freq_response}
		\phi_{mk}(f_n,C_{mk}) = 1 - \frac{2}{1 + \frac{\mathcal{D}_{mk}(f_n,C_{mk})}{\mathcal{N}_{mk}(f_n,C_{mk})}},
\end{equation}
where $\mathcal{N}_{mk}(f_n,C_{mk})$ and $\mathcal{D}_{mk}(f_n,C_{mk})$ are defined as follows:
\begin{align}
    \mathcal{N}_{mk}(f_n,C_{mk}) \!&\triangleq\! 1 \!-\! (\kappa f_n)^2(\mathpzc{L}_1 \!+\! \mathpzc{L}_2)C_{mk} \!+\! \jmath\kappa f_n R C_{mk}, \label{eqn:numerator_s} \\ 
    \mathcal{D}_{mk}(f_n,C_{mk}) \!&\triangleq\! \jmath \kappa f_n\frac{\mathpzc{L}_1}{\mathcal{Z}_0}\left(1\!-\!(\kappa f_n)^2 \mathpzc{L}_2 C_{mk}\!+\!\jmath \kappa f_n R C_{mk} \right)\!. \label{eqn:denominator_s}
\end{align}

\subsubsection{Received Signal Model} \label{Sec:Received_Model}
Every $\ell_k$-th BS-UE communication pair in the considered system model is supported by a BD-RIS-enabled wireless link: the signal emitted by the $k$-th BS is first reflected by its associated $k$-th BD-RIS and, subsequently, reaches the target $\ell_k$-th UE. For each $n$-th SC, let $\vect{H}_{k,k,n} \in \mathbb{C}^{N_{\rm ris}\times N_{\rm tx}}$ and $\vect{g}_{k,\ell_k,n} \in \mathbb{C}^{N_{\rm ris} \times 1}$ denote the channel gain matrices between the $k$-th BS and the $k$-th BD-RIS as well as between the $k$-th BD-RIS and the $\ell_k$-th UE, respectively. The vector with the reflection coefficients for each $k$-th BD-RIS at each $n$-th SC is defined as $\vect{\phi}_{k,n} \triangleq [\phi_{1k}(f_n,C_{1k}),\ldots,\phi_{Mk}(f_n,C_{Mk})]^{\rm T} \in \mathbb{C}^{N_{\rm ris} \times 1}$, and let $\vect{\Phi}_{k,n} \triangleq \operatorname{diag}\{\vect{\phi}_{k,n}\}$ $\forall k,n$. Putting all above together, the baseband received signal at the $\ell_k$-th UE on the $n$-th SC is expressed in the frequency domain as follows:
\begin{equation} \label{eqn:received_signal}
	y_{\ell_k,n} \triangleq \vect{f}_{k,\ell_k,n}^{\rm H} \vect{x}_{k,n} + \sum_{j\neq k}^K \vect{f}_{j,\ell_k,n}^{\rm H} \vect{x}_{j,k} + n_{\ell_k,n},
\end{equation}
where $n_{\ell_k,n} \sim \mathcal{CN}(0,\sigma_{\ell_k,n}^2)$ represents the AWGN, which models the thermal noises at the UE receivers. In addition, the following definitions for the channel vectors have been used:
\begin{align} 
	\vect{f}_{k,\ell_k,n}^{\rm H} &\triangleq \vect{h}_{k,\ell_k,n}^{\rm H} + \vect{g}_{k,\ell_k,n}^{\rm H}\vect{S}_k\vect{\Phi}_{k,n}\vect{H}_{k,k,n}, \label{eqn:total_channels_1}\\
	\vect{f}_{j,\ell_k,n}^{\rm H} &\triangleq \vect{h}_{j,\ell_k,n}^{\rm H} + \vect{g}_{j,\ell_k,n}^{\rm H}\vect{S}_j\vect{\Phi}_{j,n}\vect{H}_{j,j,n},
	\label{eqn:total_channels_2}
\end{align}
where each $\vect{h}_{j,\ell_k,n}\in \mathbb{C}^{N_{\rm tx} \times 1}$ indicates the direct channel at each $n$-th SC between the $\ell_k$-th UE and the $j$-th BS.

\subsection{System Design Objective and Solution} \label{Sec:Prob_Form}
Before proceeding with the design problem formulation, the following vectors are defined: \textit{i}) $\widetilde{\vect{w}} \triangleq [\tilde{\vect{w}}_1^{\rm T},\dots,\tilde{\vect{w}}_K^{\rm T}]^{\rm T}\in\mathbb{C}^{(\sum_{k=1}^{K}L_k) N_{\rm sub}N_{\rm tx}\times 1}$ with $\tilde{\vect{w}}_k \triangleq [\vect{w}_{1}^{\rm T},\dots,\vect{w}_{L_k}^{\rm T}]^{\rm T}\in\mathbb{C}^{L_k N_{\rm sub}N_{\rm tx} \times 1}$ and $\vect{w}_{\ell_k} \triangleq [\vect{w}_{\ell_k,1}^{\rm T},\ldots,\vect{w}_{\ell_k,N_{\rm sub}}^{\rm T}]^{\rm T}\in\mathbb{C}^{N_{\rm sub}N_{\rm tx}\times 1}$; and \textit{ii}) $\widetilde{\vect{c}}\triangleq [\vect{c}_1^{\rm T},\ldots,\vect{c}_K^{\rm T}]^{\rm T}\in\mathbb{R}^{KN_{\rm ris}\times 1}$ with $\vect{c}_k \triangleq [C_{1k},\ldots,C_{N_{\rm ris}k}]^{\rm T} \in \mathbb{R}^{N_{\rm ris} \times 1}$; as well as \textit{iii}) the set of matrices $\widetilde{\vect{S}} \triangleq \{\vect{S}_k\}_{k=1}^K$ including, respectively, the precoding vectors at the $K$ multi-antenna BSs, the tunable capacitances, and the switch selection matrices at the $K$ BD-RISs. Then, by treating the Multi-User Interference (MUI) term in expression~\eqref{eqn:received_signal} as an additional source of noise (specifically, colored noise), the achievable rate performance in bits/s/Hz for each $\ell_k$-th UE can be expressed as the following function of the tunable system parameter triplet $(\widetilde{\vect{w}},\widetilde{\vect{c}},\widetilde{\vect{S}})$ \cite{IBC_distributed2024_all}: 
\begin{equation} \label{eqn:sum_rate_q}	\mathcal{R}_{\ell_k}\left(\widetilde{\vect{w}},\widetilde{\vect{c}},\widetilde{\vect{S}}\right) = \frac{1}{N_{\rm sub}}\sum_{n=1}^{N_{\rm sub}} \log_2\left( 1 + \frac{|\vect{f}_{k,\ell_k,n}^{\rm H}\vect{w}_{\ell_k,n}|^2}{\operatorname{MUI}_{\ell_k,n}} \right),
\end{equation}
where each $\operatorname{MUI}_{\ell_k,n}$ is mathematically defined as follows:
\begin{align}
    \operatorname{MUI}_{\ell_k,n} &\triangleq \sigma_{\ell_k,n}^2 + \sum_{(i,j)\neq(\ell,k)} |\vect{f}_{j,\ell_k,n}^{\rm H}\vect{w}_{i_j,n}|^2 \\
    &\begin{aligned}
		=\sigma_{\ell_k,n}^2 + \underbrace{\sum_{i=1,i\neq\ell}^{L_k} |\vect{f}_{k,\ell_k,n}^{\rm H}\vect{w}_{i_k,n}|^2}_{\text{intracell interference}} + \underbrace{\sum_{j\neq k}^K \sum_{i=1}^{L_j} |\vect{f}_{j,\ell_k,n}^{\rm H}\vect{w}_{i_j,n}|^2}_{\text{intercell interference}}.
	\end{aligned}    
\end{align}
It is noted that the dependence on $\widetilde{\vect{c}}$ and $\widetilde{\vect{S}}$ is implied via the composite channels $\vect{f}_{k,\ell_k,n}$ and $\vect{f}_{j,\ell_k,n}$ as defined in \eqref{eqn:total_channels_1} and \eqref{eqn:total_channels_2}, respectively.

In what follows, the design objective for the considered RIS-enabled interference broadcast SWE is to maximize the instantaneous achievable sum-rate performance, which is mathematically formulated as follows:
\begin{align*}
	\mathcal{OP}_5: \,\max_{\widetilde{\vect{w}},\widetilde{\vect{c}},\widetilde{\vect{S}}} \, & \quad \sum_{k=1}^K \sum_{\ell=1}^{L_k} \mathcal{R}_{\ell_k}\left(\widetilde{\vect{w}},\widetilde{\vect{c}},\widetilde{\vect{S}}\right) \\
	\text{s.t.} & \quad \sum_{\ell=1}^{L_k}\sum_{n=1}^{N_{\rm sub}} \left\Vert \vect{w}_{\ell_k,n} \right\Vert^2 \leq P_k^{\max},\forall k = 1,\ldots,K, \\
    & \quad \vect{S}_k \in \mathcal{S},\forall k = 1,\ldots,K,\\
	& \quad C_{\min} \leq [\vect{c}_k]_m \leq C_{\max}, \forall m=1,\ldots,N_{\rm ris},
\end{align*}
where $\mathcal{S} \triangleq\left\{ \vect{S}\in \{0,1\}^{N_{\rm ris}\times N_{\rm ris}}:\vect{S}\vect{1} = \vect{1}, \vect{S}^{\rm T}\vect{1}=\vect{1} \right\}$ indicates the feasible set for the switch selection matrices at the BD-RISs, while $C_{\min}$ and $C_{\max}$ represent the minimum and maximum allowable values for each metasurface's tunable capacitances according to circuital characteristics, respectively. 

\subsubsection{Distributed Sum-Rate Maximization} \label{Sec:Design_SCA}
To solve $\mathcal{OP}_5$ (which is provably an NP-hard problem) in a distributed manner, the assumption that each $k$-th BS possesses the channel gain matrices included in $\vect{f}_{k,\ell_k,n}$ in \eqref{eqn:total_channels_1} $\forall n$ is necessary (see \cite{Jian2022_RIS_Hardware,HRIS_CE_all,RIS_Vahid_low,Swindlehurst_CE} for relevant CSI estimation techniques). Hereinafter, a successive concave approximation algorithmic framework is presented that enables the efficient decomposition of $\mathcal{OP}_5$ into $K$ sub-problems that can be iteratively solved in parallel by each individual BS, requiring only minimal message exchanges among their relevant processing units. In particular, let $\vect{X}_k \triangleq \{\tilde{\vect{w}}_k,\vect{c}_k,\vect{S}_k\}$ and $\vect{X}_{-k}$ be the set of all other variables except the $k$-th triplet. The sum-rate objective in $\mathcal{OP}_5$ is non-concave, due to the presence of MUI and the coupling between the design variables. Nevertheless, this function can be decomposed into the following form~\cite{scutari2013decomposition_all}:
\begin{equation} \label{eqn:total_rate}
	\overline{\mathcal{R}}(\vect{X}_k,\vect{X}_{-k})\!\triangleq\!\sum_{\ell=1}^{L_k}\mathcal{R}_{\ell_k}(\vect{X}_k,\vect{X}_{-k})+\sum_{j\neq k}^K\sum_{\ell=1}^{L_j} \mathcal{R}_{\ell_j}(\vect{X}_k,\vect{X}_{-k}).
\end{equation}
The above structure leads to the following decomposition scheme: \textit{i}) at every algorithmic iteration ${\rm t}$, the first set of terms (equal to $\mathcal{R}_k(\vect{X}_k,\vect{X}_{-k})$) is replaced by a surrogate function, denoted as $\widetilde{\mathcal{R}}_k(\vect{X}_k,\vect{X}^{\rm t})$, which depends on the current iterate $\vect{X}^{\rm t}$; and \textit{ii}) the remaining terms involved are linearized around $\vect{X}_k^{\rm t}$. To this end, the proposed updating scheme for distributively solving $\mathcal{OP}_5$ reads as: at each algorithmic iteration ${\rm t}$, each BS solves the optimization problem below:
\begin{equation*}\label{eqn:Surrogate_problem}
	\mathcal{OP}_{\mathcal{D},k}:\quad\widehat{\vect{X}}_k^{\rm t}\,=\, \arg\max_{\vect{X}_k \in \mathcal{X}_k} \, \widetilde{\mathcal{R}}_k(\vect{X}_k;\vect{X}^{\rm t})+<\boldsymbol{\Pi}^t_k,\vect{X}_k -\vect{X}_k^{\rm t}>, 
\end{equation*}
where $\mathcal{X}_k$ denotes the feasible set that combines all constraints of $\mathcal{OP}_5$, while the local surrogate function $\widetilde{\mathcal{R}}_k$ is given by:
\begin{align} \label{eqn:first_surrogate}
	\begin{split}		
		\widetilde{\mathcal{R}}_k(\vect{X}_k;\vect{X}^{\rm t}) \triangleq &\sum_{\ell=1}^{L_k}\sum_{n=1}^{N_{\rm sub}} \log_2 \left( 1 + \frac{|\vect{f}_{k,\ell_k,n}^{\rm H}\vect{w}_{\ell_k,n}|^2}{\operatorname{MUI}_{\ell_k,n}^{\rm t}} \right) +<\vect{\gamma}_{\vect{c}_k}^{\rm t},\vect{c}_k - \vect{c}_k^{\rm t}> \\
		&+ <\vect{\Gamma}_{\vect{S}_k}^{\rm t},\vect{S}_k - \vect{S}_k^{\rm t}>-\frac{\tau}{2}\Big(\Vert\vect{w}_{\ell_k} - \vect{w}_{\ell_k}^{\rm t} \Vert^2+ \Vert \vect{c}_k - \vect{c}_k^{\rm t}\Vert^2 + \Vert \vect{S}_k - \vect{S}_k^{\rm t}\Vert_{\rm F}^2\Big),
	\end{split}
\end{align}
with $\tau\!>\!0$ being an appropriately chosen parameter, $\vect{\gamma}_{\vect{c}_k}^{\rm t} \triangleq \nabla_{\vect{c}_k}\mathcal{R}_k(\vect{X}_k,\vect{X}_{-k}^{\rm t})\vert_{\vect{c}_k=\vect{c}_k^{\rm t}}$ and accordingly for $\vect{\Gamma}_{\vect{S}_k}^{\rm t}$. In addition, $\vect{\Pi}_k^{\rm t} \triangleq \sum_{j\neq k}^K \sum_{\ell=1}^{L_j} \nabla_{\vect{X}_k} \mathcal{R}_{\ell_j}(\vect{X}_k,\vect{X}_{-k}^{\rm t})\vert_{\vect{X}_k=\vect{X}_k^{\rm t}}$ which is often referred to as the pricing vector/matrix. Note that the multiplicative factor $1/N_{\rm sub}$ has been ignored in the logarithmic term, because it does not affect the optimization solution approach. Next, the solution of $\mathcal{OP}_{\mathcal{D},k}$ for each set of variables included in $\vect{X}_k$ is demonstrated. 

\subsubsection{Optimization of the Local BS Precoder} \label{Sec:Precoder}
Solving $\mathcal{OP}_5$ with respect to each BS precoder $\vect{w}_{\ell_k}$ for the $\ell_k$-th UE leads to the following optimization sub-problem:
\begin{align*}
	\begin{split}
		\mathcal{OP}_{\vect{w}_{\ell_k}}: \,\max_{\vect{w}_{\ell_k}} \, & \,\, \sum_{\ell=1}^{L_k}\Bigg(\sum_{n=1}^{N_{\rm sub}} \breve{\mathcal{R}}_{\ell_k,n}(\vect{w}_{\ell_k,n}) - \frac{\tau}{2}\Vert\vect{w}_{\ell_k} - \vect{w}_{\ell_k}^{\rm t}\Vert^2 + \Re \left\{ (\overline{\vect{\pi}}_{\ell_k}^{\rm t})^{\rm H}(\vect{w}_{\ell_k} - \vect{w}_{\ell_k}^{\rm t}) \right\}\Bigg)\\
		\text{s.t.} & \,\, \sum_{\ell=1}^{L_k}\sum_{n=1}^{N_{\rm sub}} \Vert\vect{w}_{\ell_k,n}\Vert^2 \leq P_k^{\max},
	\end{split}
\end{align*}
where $\breve{\mathcal{R}}_{\ell_k,n}(\vect{w}_{\ell_k,n})$ stands for the logarithmic term in \eqref{eqn:first_surrogate}. Also, $\overline{\vect{\pi}}_{\ell_k}^{\rm t}$ is the pricing vector associated with $\vect{w}_{\ell_k}$, which is given by $\overline{\vect{\pi}}_{\ell_k}^{\rm t} = [(\overline{\vect{\pi}}_{\ell_k,1}^{\rm t})^T,\ldots,(\overline{\vect{\pi}}_{\ell_k,N_{\rm sub}}^{\rm t})^T]^T\in\mathbb{C}^{N_{\rm sub}N_{\rm tx}\times 1}$ with
\begin{equation} \label{eqn:pricing_w_q}
		\overline{\vect{\pi}}_{\ell_k,n}^{\rm t} \!=\! -\frac{1}{\ln(2)}\sum_{j \neq k}^K \sum_{i=1}^{L_j} \frac{\operatorname{snr}_{i_j,n}^{\rm t}}{(1 + \operatorname{snr}_{i_j,n}^{\rm t})\operatorname{MUI}_{i_j,n}^{\rm t}}\vect{f}_{k,i_j,n} \vect{f}_{k,i_j,k}^{\rm H} \vect{w}_{\ell_k,n}^{\rm t},
\end{equation}
where the factor $\operatorname{snr}_{i_j,n}^{\rm t}$ is defined as $\operatorname{snr}_{i_j,n}^{\rm t} \triangleq \left\lvert \vect{f}_{j,i_j,k}^{\rm H}\vect{w}_{i_j,k}^{\rm t} \right\rvert^2/\operatorname{MUI}_{i_j,n}^{\rm t}$. The optimization problem $\mathcal{OP}_{\vect{w}_{\ell_k}}$ remains intrinsically non-concave, mainly because $\breve{\mathcal{R}}_{\ell_k,n}(\vect{w}_{\ell_k,n})$ contains a logarithmic term applied to a quadratic expression in $\vect{w}_{\ell_k,n}$. This structure prevents direct application of standard convex optimization techniques. This difficulty can be tackled by employing the following surrogate function:
\begin{equation} \label{eqn:logarithmic_surrogate}
		\begin{aligned}
			\widehat{\mathcal{R}}_{\ell_k,n} \!=\! -a_{\ell_k,n}^{\rm t}\vect{w}_{\ell_k,n}^{\rm H} \vect{F}_{\ell_k,n} \vect{w}_{\ell_k,n} + 2\Re\{(\vect{b}_{\ell_k,n}^{\rm t})^{\rm H}\vect{w}_{\ell_k,n}\},
		\end{aligned}
\end{equation}
where $\vect{F}_{\ell_k,n} \triangleq \vect{f}_{k,\ell_k,n}\vect{f}_{k,\ell_k,n}^{\rm H}$, and $a_{\ell_k,n}^{\rm t}$ and $\vect{b}_{\ell_k,n}^{\rm t}$ are defined as follows:
\begin{align}
    a_{\ell_k,n}^{\rm t} &\triangleq \frac{1}{\ln(2)}\frac{\lvert \vect{f}_{k,\ell_k,n}^{\rm H} \vect{w}_{\ell_k,n}^{\rm t} \rvert^2}{\left(\operatorname{MUI}_{\ell_k,n}^{\rm t} + \lvert \vect{f}_{k,\ell_k,n}^{\rm H} \vect{w}_{\ell_k,n}^{\rm t} \rvert^2\right) \operatorname{MUI}_{\ell_k,n}^{\rm t}}, \label{eqn:surrog_a} \\
    \vect{b}_{\ell_k,n}^{\rm t} &\triangleq \frac{1}{\ln(2)}\frac{1}{\operatorname{MUI}_{\ell_k,n}^{\rm t}} \vect{F}_{\ell_k,n} \vect{w}_{\ell_k,n}^{\rm t}. \label{eqn:surrog_b}
\end{align}

Next, by defining the block diagonal matrix $\tilde{\vect{F}}_{\ell_k} \triangleq \operatorname{blkdiag}\{a_{\ell_k,n}^{\rm t}\vect{F}_{\ell_k,n}\}_{n=1}^{N_{\rm sub}}$, and the vector $\tilde{\vect{f}}_{\ell_k} \triangleq [(\vect{b}_{\ell_k,1}^{\rm t})^{\rm T},\ldots,(\vect{b}_{\ell_k,N_{\rm sub}}^{\rm t})^{\rm T}]^{\rm T}$, $\mathcal{OP}_{\vect{w}_{\ell_k}}$'s objective function becomes:
\begin{equation} \label{eqn:compact_surrogate_w}
	\mathcal{J}=-\vect{w}_{\ell_k}^{\rm H}\left(\tilde{\vect{F}}_{\ell_k} + \frac{\tau}{2}\vect{I}_{N_{\rm sub}N_{\rm tx}} \right)\vect{w}_{\ell_k} + \Re\left\{(\vect{v}_{\ell_k}^{\rm t})^{\rm H}\vect{w}_{\ell_k} \right\},
\end{equation}
where $\vect{v}_{\ell_k}^{\rm t} \triangleq \overline{\vect{\pi}}_{\ell_k}^{\rm t} + 2\tilde{\vect{f}}_{\ell_k} + \tau \vect{w}_{\ell_k}^{\rm t}$. It can be trivially shown that $\tilde{\vect{F}}_{\ell_k}\succeq\vect{0}$ yielding the concavity of \eqref{eqn:compact_surrogate_w}. Therefore, the optimal $\vect{w}_{\ell_k}$ ($\vect{w}_{\ell_k}^{\rm opt}$) follows by the first-order condition, which results in:
\begin{equation} \label{eqn:optimal_w}
	\vect{w}_{\ell_k}^{\rm opt}(\lambda) = \left(\tilde{\vect{F}}_{\ell_k} + \left(\frac{\tau}{2} + \lambda\right)\vect{I}_{N_{\rm sub}N_{\rm tx}} \right)^{-1} \vect{v}_{\ell_k}^{\rm t},
\end{equation}
where $\lambda \geq 0$ denotes the Lagrange multiplier associated with the transmit power constraint in $\mathcal{OP}_{\vect{w}_{\ell_k}}$, whose optimum value ($\lambda^{\rm opt}$) can be obtained by elaborating on Slater's condition and a bisection search.

\subsubsection{Optimization of the Local RIS Phase Configuration} \label{Sec:RIS_Config}
For every $n$-th SC and each $k$-th BD-RIS, the corresponding phase configuration vector $\vect{\phi}_{k,n}$ depends functionally on the set of design parameters collected in $\vect{c}_k$. This dependence allows this parameter to be adjusted by formulating and solving the following reduced optimization problem $\mathcal{OP}_{\vect{c}_k}$, in which the entries of $\vect{c}_k$ serve as the underlying decision variables that dictate the resulting reflection profile: 
\begin{align*}
	\mathcal{OP}_{\vect{c}_k}: \,\max_{\vect{c}_k} \, & \,\, - \frac{\tau}{2}\Vert\vect{c}_k - \vect{c}_k^{\rm t}\Vert^2 + \Re\left\{ (\vect{\gamma}_{\vect{c}_k}^{\rm t} + \underline{\vect{\pi}}_k^{\rm t})^{\rm H}(\vect{c}_k - \vect{c}_k^{\rm t})\right\} \\
	\text{s.t.} & \quad C_{\min} \leq [\vect{c}_k]_m \leq C_{\max} \, \, \forall m=1,\ldots,N_{\rm RIS},
\end{align*}
which is clearly a concave optimization problem. Before proceeding to this problem's solution, analytic expressions for $\vect{\gamma}_{\vect{c}_k}^{\rm t}$ and $\underline{\vect{\pi}}_k^{\rm t}$ are derived in the sequel using the following matrix definitions:
\begin{align}
    \vect{A}_{k,\ell_k,n} &\triangleq \vect{H}_{k,k,n}\vect{w}_{\ell_k,n}\vect{w}_{\ell_k,n}^{\rm H}\vect{h}_{k,\ell_k,n}\vect{g}_{k,\ell_k,n}^{\rm H}\vect{S}_k, \\
    \vect{A}_{k,\ell_k,n}^{m_k} &\triangleq \vect{H}_{k,k,n}\vect{w}_{m_k,n}\vect{w}_{m_k,n}^{\rm H}\vect{h}_{k,\ell_k,n}\vect{g}_{k,\ell_k,n}^{\rm H}\vect{S}_k, \\
    \vect{A}_{k,i_j,n} &\triangleq \vect{H}_{k,k,n}\left(\sum_{\ell=1}^{L_k}\vect{w}_{\ell_k,n}\vect{w}_{\ell_k,n}^{\rm H}\right)\vect{h}_{k,i_j,n}\vect{g}_{k,i_j,n}^{\rm H}\vect{S}_k, \\
    \vect{B}_{k,\ell_k,n} &\triangleq \vect{S}_k^{\rm T}\vect{g}_{k,\ell_k,n}\vect{g}_{k,\ell_k,n}^{\rm H}\vect{S}_k, \\
    \vect{B}_{k,i_j,n} &\triangleq \vect{S}_k^{\rm T}\vect{g}_{k,i_j,n}\vect{g}_{k,i_j,n}^{\rm H}\vect{S}_k, \\
    \vect{C}_{k,\ell_k,n} &\triangleq \vect{H}_{k,k,n}\vect{w}_{\ell_k,n}\vect{w}_{\ell_k,n}^{\rm H}\vect{H}_{k,k,n}^{\rm H}, \\
    \vect{C}_{k,m_k,n} &\triangleq \vect{H}_{k,k,n}\vect{w}_{m_k,n}\vect{w}_{m_k,n}^{\rm H}\vect{H}_{k,k,n}^{\rm H}, \\
    \vect{M}_{k,\ell_k,n} &\triangleq \vect{A}_{k,\ell_k,n} + \vect{C}_{k,\ell_k,n} (\vect{\Phi}_{k,n}^{\rm t})^{\rm H} \vect{B}_{k,\ell_k,n}, \\
    \vect{M}_{k,\ell_k,n}^{m_k} &\triangleq \vect{A}_{k,\ell_k,n}^{m_k} + \vect{C}_{k,m_k,n} (\vect{\Phi}_{k,n}^{\rm t})^{\rm H} \vect{B}_{k,\ell_k,n}, \\
    \vect{M}_{k,i_j,n} &\triangleq \vect{A}_{k,i_j,n} + \left(\sum_{\ell=1}^{L_k}\vect{C}_{k,\ell_k,n}\right) (\vect{\Phi}_{k,n}^{\rm t})^{\rm H} \vect{B}_{k,i_j,n}, \\
    \vect{Q}_{k,n} &\triangleq \textrm{diag}\left\{\frac{\partial([\vect{\phi}_{kn}]_1)^*}{\partial C_{1m}},\ldots,\frac{\partial([\vect{\phi}_{kn}]_{N_{\rm ris}})^*}{\partial C_{N_{\rm ris}k}}\right\},
\end{align}
where the partial derivatives of $[\vect{\phi}_{k,n}]_m$ $\forall$$k,n,m$ with respect to the BD-RIS tunable capacitance $C_{mk}$ can be computed as follows:
\begin{align} \label{eqn:der_C_m}  
    \nonumber\frac{\partial([\vect{\phi}_{k,n}]_m)^*}{\partial C_{mk}} =&\frac{-2}{\left(\mathcal{N}_{mk}^*(f_n,C_{mk}) + \mathcal{D}_{mk}^*(f_n,C_{mk}) \right)^2}\\
    &\times\Bigg( \frac{\partial \mathcal{N}_{mk}^*(f_n,C_{mk})}{\partial C_{mk}} \mathcal{D}_{mk}^*(f_n,C_{mk})- \mathcal{N}_{mk}^*(f_n,C_{mk}) \frac{\partial \mathcal{D}_{mk}^*(f_n,C_{mk})}{\partial C_{mk}} \Bigg),\nonumber
\end{align}
where respectively following \eqref{eqn:numerator_s} and \eqref{eqn:denominator_s} holds that: 
\begin{align}
    \frac{\partial \mathcal{N}_{mk}^*(f_n,C_{mk})}{\partial C_{mk}} &= -\left(\kappa f_n\right)^2\left(\mathpzc{L}_1 + \mathpzc{L}_2\right) - \jmath \kappa f_n R, \\
    \frac{\partial \mathcal{D}_{mk}^*(f_n,C_{mk})}{\partial C_{mk}} &= -\jmath \kappa f_n \frac{\mathpzc{L}_1}{\mathcal{Z}_0}\left(-\left(\kappa f_n\right)^2 \mathpzc{L}_2 - \jmath \kappa f_n R\right).
\end{align}
Then, the vectors $\vect{\gamma}_{\vect{c}_{k}}^{\rm t}$ and $\underline{\vect{\pi}}_{\vect{c}_{k}}^{\rm t}$ in $\mathcal{OP}_{\vect{c}_k}$ are given by the following analytic expressions:
\begin{align}
    &\begin{aligned}
        &\vect{\gamma}_{\vect{c}_k}^{\rm t} = \sum_{\ell=1}^{L_k}\sum_{n=1}^{N_{\rm sub}} \frac{2/\ln(2)}{(1+\operatorname{snr}_{\ell_k,n}^{\rm t})(\operatorname{MUI}_{\ell_k,n}^{\rm t})^2} \\
        &\times \Bigg( \Re\Bigg\{ \operatorname{MUI}_{\ell_k,n}^{\rm t} \vect{Q}_{k,n} \operatorname{vec}_{\rm d}(\vect{M}_{k,\ell_k,n}) \Bigg\} - |\vect{f}_{k,\ell_k,n}^{\rm H}\vect{w}_{\ell_k,n}|^2 \sum_{m\neq\ell}^{L_k}\Re\Bigg\{\vect{Q}_{k,n}\operatorname{vec}_{\rm d}(\vect{M}_{k,\ell_k,n}^{m_k}) \Bigg\} \Bigg), 
    \end{aligned} \label{eqn:gamma_RIS}\\
    &\underline{\vect{\pi}}_{\vect{c}_{k}}^{\rm t} = -\frac{2}{\ln(2)}\sum_{j \neq k}^K \sum_{i=1}^{L_j} \sum_{n=1}^{N_{\rm sub}}\frac{\operatorname{snr}_{i_j,k}^{\rm t}}{(1+\operatorname{snr}_{i_j,k}^{\rm t})\operatorname{MUI}_{i_j,k}^{\rm t}} \Re\Bigg\{\vect{Q}_{k,n}\operatorname{vec}_{\rm d}(\vect{M}_{k,i_j,n})\Bigg\}. \label{eqn:pricing_RIS}
\end{align}

Now, $\mathcal{OP}_{\vect{c}_k}$ can be solved in closed form. In particular, by letting $\vect{\beta}_k \triangleq \tau\vect{c}_k^{\rm t} + \vect{\gamma}_{\vect{c}_k}^{\rm t} + \underline{\vect{\pi}}_{\vect{c}_{k}}^{\rm t}$, $\mathcal{OP}_{\vect{c}_k}$'s optimal solution is given by:
\begin{equation} \label{eqn:RIS_solution}
    [\vect{c}_k]_m^{\rm opt} =  \begin{cases}
        C_{\min}, & \text{if}\,\,\frac{1}{\tau}[\vect{\beta}_k]_m < C_{\min} \\
        C_{\max}, & \text{if}\,\,\frac{1}{\tau}[\vect{\beta}_k]_m >C_{\max} \\
        \frac{1}{\tau}[\vect{\beta}_k]_m, & \text{otherwise}
    \end{cases}.
\end{equation}

\subsubsection{Optimization of the Local RIS Switch Selection Matrix} \label{Sec:Non_Diag_Design}
The design of the switch selection matrix $\vect{S}_k$ at each $k$-th BD-RIS reduces to the following simplified optimization problem (it is noted that $\textrm{Tr}(\vect{S}_k\vect{S}_k^{\rm T}) = N_{\rm ris}$):
\begin{align*}
	\mathcal{OP}_{\vect{S}_k}: \max_{\vect{S}_k} \, & \,\, \textrm{Tr}\left( \Re\left\{\vect{\Gamma}_{\vect{S}_k}^{\rm t} + \vect{\Pi}_{\vect{S}_k}^{\rm t} + \tau\vect{S}_k^{\rm t}\right\}^{\rm H}\vect{S}_k \right) \\
	\text{s.t.} & \quad [\vect{S}_k]_{i,j}\in\{0,1\},\forall i,j = 1,\ldots,N_{\rm ris}, \\
	& \quad \sum_{i=1}^{N_{\rm ris}}[\vect{S}_k]_{i,j} = 1, \forall j = 1,\ldots,N_{\rm ris}, \\
	& \quad \sum_{j=1}^{N_{\rm ris}}[\vect{S}_k]_{i,j} = 1, \forall i = 1,\ldots,N_{\rm ris},
\end{align*}
whose solution depends on $\vect{\Gamma}_{\vect{S}_k}^{\rm t}$ and $\vect{\Pi}_{\vect{S}_k}^{\rm t}$ derived below. Let first the following matrix definitions:
\begin{align} 
    \vect{F}_{k,\ell_k,n} &\triangleq \vect{\Phi}_{k,n}\vect{H}_{k,k,n}\vect{w}_{\ell_k,n}\vect{w}_{\ell_k,n}^{\rm H}\vect{h}_{k,\ell_k,n}\vect{g}_{k,\ell_k,n}^{\rm H}, \\
    \vect{F}_{k,\ell_k,n}^{m_k} &\triangleq \vect{\Phi}_{k,n}\vect{H}_{k,k,n}\vect{w}_{m_k,n}\vect{w}_{m_k,n}^{\rm H}\vect{h}_{k,\ell_k,n}\vect{g}_{k,\ell_k,n}^{\rm H}, \\
    \vect{F}_{k,i_j,n} &\triangleq \vect{\Phi}_{k,n}\vect{H}_{k,k,n}\left(\sum_{\ell=1}^{L_k}\vect{w}_{\ell_k,n}\vect{w}_{\ell_k,n}^{\rm H}\right)\vect{h}_{k,i_j,n}\vect{g}_{k,i_j,n}^{\rm H},  \\
    \vect{K}_{k,\ell_k,n} &\triangleq \vect{\Phi}_{k,n}\vect{H}_{k,k,n}\vect{w}_{\ell_k,n}\vect{w}_{\ell_k,n}^{\rm H}\vect{H}_{k,k,n}^{\rm H}\vect{\Phi}_{k,n}^{\rm H},  \\
    \vect{K}_{k,\ell_k,n}^{m_k} &\triangleq \vect{\Phi}_{k,n}\vect{H}_{k,k,n}\vect{w}_{m_k,n}\vect{w}_{m_k,n}^{\rm H}\vect{H}_{k,k,n}^{\rm H}\vect{\Phi}_{k,n}^{\rm H},  \\
    \vect{G}_{k,\ell_k,n} &\triangleq \vect{g}_{k,\ell_k,n}\vect{g}_{k,\ell_k,n}^{\rm H}, \,\,\vect{G}_{k,n_k,n} \triangleq \vect{g}_{k,i_j,n}\vect{g}_{k,i_j,n}^{\rm H}, \\
    \vect{N}_{k,\ell_k,n} &\triangleq \vect{F}_{k,\ell_k,n} + \vect{K}_{k,\ell_k,n} (\vect{S}_k^{\rm t})^{\rm T} \vect{G}_{k,\ell_k,n}, \\
    \vect{N}_{k,\ell_k,n}^{m_k} &\triangleq \vect{F}_{k,\ell_k,n}^{m_k} + \vect{K}_{k,\ell_k,n}^{m_k} (\vect{S}_k^{\rm t})^{\rm T} \vect{G}_{k,\ell_k,n}, \\
    \vect{N}_{k,i_j,n} &\triangleq \vect{F}_{k,i_j,n} + \left(\sum_{\ell=1}^{L_k}\vect{K}_{k,\ell_k,n}\right) (\vect{S}_k^{\rm t})^{\rm T} \vect{G}_{k,i_j,n}.
\end{align}
Then, $\vect{\Gamma}_{\vect{S}_k}^{\rm t}$ and $\vect{\Pi}_{\vect{S}_k}^{\rm t}$ are given by:
\begin{align}
    &\begin{aligned}
        \vect{\Gamma}_{\vect{S}_k}^{\rm t} =& \frac{2}{\ln(2)}\sum_{\ell=1}^{L_k}\sum_{n=1}^{N_{\rm sub}} \frac{1}{(1+\operatorname{snr}_{\ell_k,n}^{\rm t})(\operatorname{MUI}_{\ell_k,n}^{\rm t})^2}\\
        &\times \Bigg(\!\operatorname{MUI}_{\ell_k,n}^{\rm t}\vect{N}_{k,\ell_k,n}-|\vect{f}_{k,\ell_k,n}^{\rm H}\vect{w}_{\ell_k,n}|^2\sum_{m\neq\ell}^{L_k}\vect{N}_{k,\ell_k,n}^{m_k}\!\Bigg)^{\rm T}, \\
    \end{aligned}  \label{eqn:gamma_Sel_Mat}\\
    &\begin{aligned} 
        \vect{\Pi}_{\vect{S}_k}^{\rm t} = &-\frac{2}{\ln(2)}\sum_{j\neq k}^K \sum_{i=1}^{L_j} \sum_{n=1}^{N_{\rm sub}}\frac{\operatorname{snr}_{i_j,n}^{\rm t}}{(1+\operatorname{snr}_{i_j,n}^{\rm t})\operatorname{MUI}_{i_j,n}^{\rm t}} \vect{N}_{k,i_j,k}^{\rm T}.
    \end{aligned} \label{eqn:pricing_Sel_Mat}
\end{align}

\begin{algorithm}[!t]
	\begin{algorithmic}[1]
		\caption{Solution of $\mathcal{OP}_5$}
		\label{alg:OP_Overall_Distributed_Algorithm}
		\State \textbf{Input:} ${\rm t}=0$, $\{\alpha^{\rm t}\}\geq 0$, $\tau>0$, $\epsilon > 0$, $K$, as well as feasible $\widetilde{\vect{w}}^{(0)}$, $\widetilde{\vect{c}}^{(0)}$, $\widetilde{\vect{S}}^{(0)}$, and $\overline{\mathcal{R}}^{(0)}$ as defined in \eqref{eqn:total_rate}.
		\State Compute $\vect{\phi}_{k,n}^{(0)}\,\,\forall k,n$ as a function of $\widetilde{\vect{c}}^{(0)}$ using \eqref{eqn:RIS_freq_response}.
		\For{$ {\rm t} = 1,2,\ldots$}
		\For{$ k = 1,2,\ldots,K$}
		\State 
		\parbox[t]{\dimexpr\linewidth-\algorithmicindent}{%
			Compute $\vect{f}_{k,\ell_k,n}$ and $\vect{f}_{j,\ell_k,n}\,\,\forall j\neq k$ according to \eqref{eqn:total_channels_1} and \eqref{eqn:total_channels_2}.
		}
		\State Compute the pricing vector $\overline{\vect{\pi}}_{\ell_k,n}^{\rm t}$ using \eqref{eqn:pricing_w_q}.
		\State Compute $a_{\ell_k,n}^{\rm t}$ using \eqref{eqn:surrog_a} and $\vect{b}_{\ell_k,n}^{\rm t}$ using \eqref{eqn:surrog_b}.
		\State 
		\parbox[t]{\dimexpr\linewidth-\algorithmicindent}{%
			Formulate the block diagonal matrix $\tilde{\vect{F}}_{\ell_k}$ and compute the vectors:\\ $\tilde{\vect{f}}_{\ell_k} = [(\vect{b}_{\ell_k,1}^{\rm t})^{\rm T},\ldots,(\vect{b}_{\ell_k,N_{\rm sub}}^{\rm t})^{\rm T}]^{\rm T}$ and $\vect{v}_{\ell_k}^{\rm t} = \overline{\vect{\pi}}_{\ell_k}^{\rm t} + 2\tilde{\vect{f}}_{\ell_k} + \tau \vect{w}_{\ell_k}^{{\rm t}-1}$.
		}
		\State 
		\parbox[t]{\dimexpr\linewidth-\algorithmicindent}{%
			Compute $\widehat{\vect{w}}_{\ell_k}^{\rm t}$ according to \eqref{eqn:optimal_w} and a bisection search method.
		}
		\State Compute $\vect{\gamma}_{\vect{c}_k}^{\rm t}$ and $\underline{\vect{\pi}}_{\vect{c}_k}^{\rm t}$ according to \eqref{eqn:gamma_RIS} and \eqref{eqn:pricing_RIS}, respectively.
		\State Compute $\widehat{\vect{c}}_k^{\rm t}$ according to \eqref{eqn:RIS_solution}.
		\State Compute $\vect{\Gamma}_{\vect{S}_k}^{\rm t}$ and $\vect{\Pi}_{\vect{S}_k}^{\rm t}$ according to \eqref{eqn:gamma_Sel_Mat} and \eqref{eqn:pricing_Sel_Mat}, respectively.
		\State 
		\parbox[t]{\dimexpr\linewidth-\algorithmicindent}{%
			Solve the linear assignment problem $\mathcal{OP}_{\vect{S}_k}$ numerically to compute $\widehat{\vect{S}}_k^{\rm t}$.
		}
        \State Collect $\vect{\Pi}_k^{\rm t} = \{\overline{\vect{\pi}}_{kn}^{\rm t},\underline{\vect{\pi}}_{\vect{c}_k}^{\rm t},\vect{\Pi}_{\vect{S}_k}^{\rm t}\}$ and transmit to UEs $j\neq k$.
		\State Obtain $\widehat{\vect{X}}_k^{\rm t} = \left\{\widehat{\vect{w}}_k^{\rm t},\widehat{\vect{c}}_k^{\rm t},\widehat{\vect{S}}_k^{\rm t}\right\}$ and $\vect{X}_k^{{\rm t}+1} = \vect{X}_k^{{\rm t}} + \alpha^{\rm t}\left(\widehat{\vect{X}}_k^{\rm t} - \vect{X}_k^{{\rm t}}\right)$.
		\EndFor
		\If $\left\lvert\left(\overline{\mathcal{R}}^{({\rm t})} - \overline{\mathcal{R}}^{({\rm t}-1)}\right)/\overline{\mathcal{R}}^{({\rm t})}\right\rvert \leq \epsilon$, \textbf{break}; 
		\EndIf
		\EndFor
		\State \textbf{Output:} $\widetilde{\vect{w}}^{({\rm t})}$, $\widetilde{\vect{c}}^{({\rm t})}$, and $\widetilde{\vect{S}}^{({\rm t})}$.
	\end{algorithmic}
\end{algorithm}

With the above expressions for $\vect{\Gamma}_{\vect{S}_k}^{\rm t}$ and $\vect{\Pi}_{\vect{S}_k}^{\rm t}$, $\mathcal{OP}_{\vect{S}_k}$ can be tackled, without loss of optimality, by dropping the binary constraints and relaxing the rest of them \cite{burkard2012assignment}. Then, it can be efficiently solved as a linear program. 

The overall algorithmic steps proposed to tackle $\mathcal{OP}_5$ are presented in Algorithm~\ref{alg:OP_Overall_Distributed_Algorithm}. 

\subsection{Numerical Results and Discussion} \label{Sec:Perf_Eval}
In this section, performance evaluation results for the distributed design described in Section~\ref{Sec:Prob_Form} for the considered wideband BD-RIS-empowered interference MISO broadcast system are presented. Specifically, to design the precoding vectors at the $K$ BSs as well as the tunable capacitances and switch selection matrices at the $K$ BD-RISs, Algorithm~\ref{alg:OP_Overall_Distributed_Algorithm} has been used. The results have been obtained by numerically evaluating the achievable sum-rate performance metric given in expression~\eqref{eqn:total_rate}. 

In the conducted simulations, all nodes were considered positioned on a Cartesian system with coordinates given by the triad $(x,y,h)$, where $x$ and $y$ denote the coordinates on the $x$- and $y$-axis, respectively, while $h$ represents the node's height, i.e., its positive value on the $z$-axis. $K = 4$ BSs were assumed located in a rectangle of width $w = 60$~m and length $r = 120$~m, with BS $1$ placed at the origin in height $h_{\rm BS_1}$, BS $2$ at the location $(w,0,h_{\rm BS_2})$, BS $3$ at $(0,r,h_{\rm BS_3})$, and BS $4$ at $(w,r,h_{\rm BS_4})$, with $h_{\rm BS_k} = 5$~m $\forall k=1,2,3$, and $4$. Moreover, it was further assumed that there are a total of $14$ UEs, organized in circular clusters with respect to each deployed BS, as follows: $L_1 = 2$, $L_2 = 3$, $L_3 = 4$, and $L_4 = 5$. In particular, all $K=4$ clusters shared the same radius, equal to $r_{\rm cl}=3$~m, their centers were placed on the $xy$-plane at the coordinates $(20,60)$, $(40,60)$, $(25,60)$, and $(35,60)$, respectively, while the exact position of each UE was randomly generated so that it belongs to its specific cluster. In addition, the coordinates of the $K = 4$ BD-RISs on the $xy$-plane were fixed to $(22.5,63.75)$ for BD-RIS $1$, $(37.5,63.75)$ for BD-RIS $2$, $(22.5,56.25)$ for BD-RIS $3$, and $(37.5,56.25)$ for BD-RIS $4$, while all shared the same $z$-coordinate value $h_{\rm BD-RIS}= 3$~m. All wireless channels were modeled as wideband fading channels with $D$ delay taps in their time-domain impulse responses, where each tap coefficient was modeled as a circularly symmetric complex Gaussian random variable. To obtain the corresponding frequency-domain representation, the block-cyclic matrices $\widetilde{\vect{H}}_{kk} \in \mathbb{C}^{N_{\rm sub}\times N_{\rm sub}N_{\rm tx}}$, $\widehat{\vect{H}}_{kk} \in \mathbb{C}^{N_{\rm ris}N_{\rm sub}\times N_{\rm sub}N_{\rm tx}}$ and $\widetilde{\vect{G}}_{kk} \in \mathbb{C}^{N_{\rm sub}\times N_{\rm sub}N_{\rm ris}}$ were first built. The first block of $\widehat{\vect{H}}_{kk}$ is given by $\left[ (\widetilde{\vect{h}}_{kk,0})^{\rm H},\dots,(\widetilde{\vect{h}}_{kk,D-1})^{\rm H},\vect{0}_{N_{\rm tx}}^{\rm T},\ldots,\vect{0}_{N_{\rm tx}}^{\rm T} \right]^{\rm H}$, and accordingly for $\widehat{\vect{H}}_{kk}$ and $\widetilde{\vect{G}}_{kk}$. In these expression, $\widetilde{\vect{h}}_{kk,d}$ denotes the impulse response coefficients of the corresponding channels at the $d$-th delay tap ($d = 0,1,\dots,D-1$). Hence, these matrices can be organized as sequences of cyclic matrices, enabling the DFT application (via the normalized DFT matrix $\vect{F}_{\rm DFT}$) to represent the channels in the frequency domain.

Distance-dependent pathloss between any two nodes $i$ and $j$ separated by a distance $d_{i,j}$, where $i$ and $j$ are either a BS, a BD-RIS, or a UE, was also considered. Specifically, the pathloss was modeled as ${\rm PL}_{i,j} \triangleq {\rm PL_0}(d_{i,j}/d_0)^{-\alpha_{i,j}}$ with ${\rm PL_0} \triangleq (\frac{\lambda_c}{4\pi})^2$ representing the signal attenuation at the reference distance $d_0 = 1$~m, where $\lambda_c$ denotes the carrier wavelength. The distance $d_{i,j}$ was evaluated for each $k$-th BS antenna element and each $k$-th UE, each $k$-th BS antenna element and each $k$-th BD-RIS element, as well as each $k$-th BD-RIS element and each $k$-th UE, in order to determine the pathloss of the channels $\vect{h}_{kk,n}$, $\vect{H}_{kk,n}$, and $\vect{g}_{kk,n}$, respectively. The pathloss exponents were chosen as $\alpha_{\rm BS,UE} = 3.7$, $\alpha_{\rm BS,BD-RIS} = 2.6$, and $\alpha_{\rm BD-RIS,UE} = 2.2$. In addition, it was assumed that each BS employs a uniform linear array and each BD-RIS uses a uniform planar array, both placed on the $xz$-plane and with inter-element spacing equal to $\lambda_c/2$. 

In the subsequent performance evaluations, an equal transmit power budget at all BSs and identical noise variances at all UEs were applied, setting specifically $P_k = P_{\max}$ and $\sigma_k^2 = \sigma^2 = -80$ dBm $\forall$~$k = 1,2,3,4$. The carrier frequency was chosen as $f_c = 2.4$~GHz, the bandwidth as ${\rm BW} = 100$ MHz, and the number of SCs was set as $N_{\rm sub} = 16$. The delay spread was represented with $D = 16$ taps, and the cyclic prefix length was set to $N_{\rm cp} = 16$. The circuit parameters for each BD-RIS were set as follows: $\mathpzc{L}_1 = 2.5$ nH, $\mathpzc{L}_2 = 0.7$ nH, $R = 1$ ${\rm \Omega}$, free-space impedance $\mathcal{Z}_0 = 50$ $\Omega$, $C_{\min} = 0.2$ pF, and $C_{\max} = 3$ pF. Finally, for the algorithmic step size, the time-varying rule $\alpha^{\rm t} = \frac{\alpha^{{\rm t}-1}+ a({\rm t})}{1+b({\rm t})}$ (with $a({\rm t}) = a = 0.9$ and $b({\rm t})=b{\rm t}$ where $b = 0.95$) at each ${\rm t}$-th algorithmic iteration was adopted, along with the initialization $\alpha^0 = 1$ for the updates of the BS precoders $\vect{w}_k$'s and the BD-RIS parameters $\vect{c}_k$'s. In contrast, for the selection matrices $\vect{S}_k$'s, $\alpha^{\rm t} = \alpha = 1$ was fixed for all ${\rm t}$ values to ensure that the constraints associated with variables are not violated. All performance evaluation results reported hereinafter have been obtained using $100$ independent MC channel realizations.

\begin{figure}[!t]
	\centering
	\includegraphics[width=3.45in]{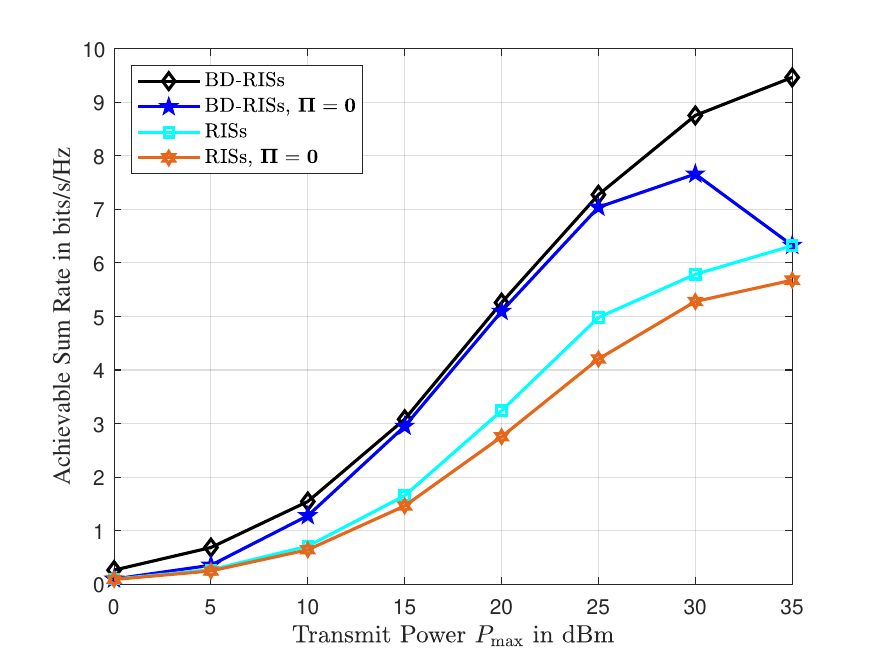}
	\caption{Achievable sum-rate performance in bits/s/Hz versus the transmit power $P_{\max}$ for every of the $K=2$ BSs each equipped with $N_{\rm tx}=8$ antennas, considering $K=2$ frequency-selective BD-RISs or conventional diagonal RISs each with $N_{\rm ris}=144$ unit elements (i.e., $\vect{S}_1 = \vect{S}_2 = \mathbf{I}_{144}$). All frequency-selective wireless fading channels were simulated as Rayleigh distributed considering $N_{\rm sub}=16$ SCs at ${\rm BW}=100$ MHz of bandwidth on the carrier frequency $f_c=2.4$ GHz. The performance of cooperative and non-cooperative ($\vect{\Pi}=\vect{0}$) beamforming schemes using Algorithm~\ref{alg:OP_Overall_Distributed_Algorithm} is compared.}
	\label{fig:Rates_vs_TX_Power}
\end{figure}

The achievable sum-rate performance is plotted in Fig.~\ref{fig:Rates_vs_TX_Power} as a function of the maximum transmit power for every of the $K=2$ BSs, each equipped with $N_{\rm tx}=8$ antennas. Frequency-selective BD-RISs and conventional diagonal RISs each with $N_{\rm ris}=144$ unit elements (i.e., $\vect{S}_1 = \vect{S}_2 = \mathbf{I}_{144}$) were considered, whereas $L_1=2$ and $L_2=3$ UEs were assumed for the first (served by BS $1$) and the second (served by BS $2$) cell, respectively. The sum-rate performance of cooperative and non-cooperative ($\vect{\Pi}=\vect{0}$) beamforming schemes has been simulated using the proposed Algorithm~\ref{alg:OP_Overall_Distributed_Algorithm}. It can be observed that, as $P_{\max}$ increases, both cooperative and non-cooperative schemes admit higher rates when BD-RISs are deployed, a fact that is attributed to the additional degrees of freedom offered by this architecture in comparison with diagonal RISs. In addition, it is depicted that, across the entire range of $P_{\max}$, the presented BS cooperation approach outperforms the non-cooperative case. 

\begin{figure}[!t]
	\centering
	\includegraphics[width=3.45in]{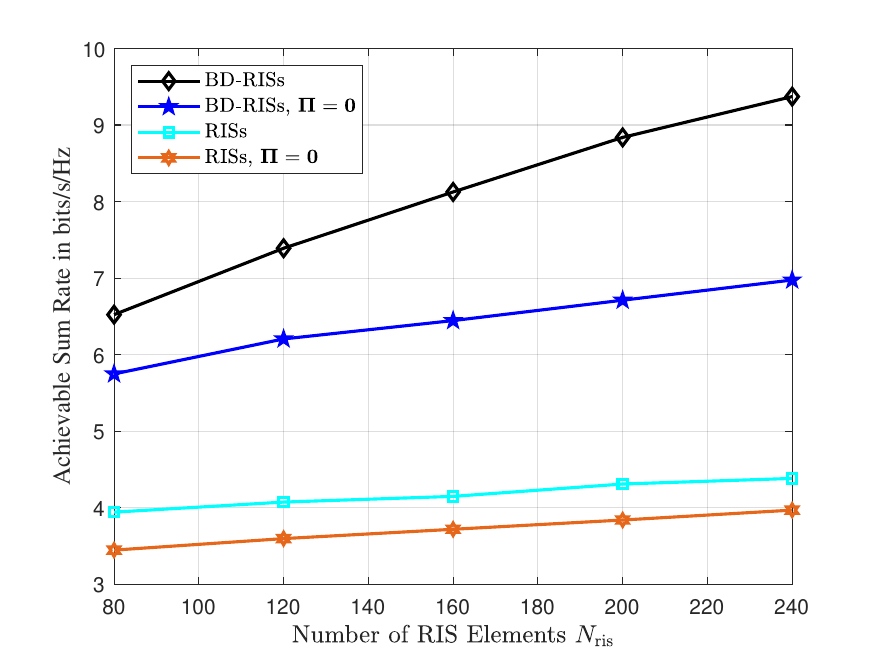}
	\caption{Achievable sum-rate performance in bits/s/Hz versus the common number $N_{\rm ris}$ of unit elements at each of the $K=2$ frequency-selective BD-RISs or conventional diagonal RISs (i.e., $\vect{S}_1 = \vect{S}_2 = \mathbf{I}_{N_{\rm ris}}$) for both cooperative and non-cooperative ($\vect{\Pi}=\vect{0}$) beamforming schemes using Algorithm~\ref{alg:OP_Overall_Distributed_Algorithm}, considering the transmit power $P_{\max}=30$~dBm for every of the $K=2$ BSs each equipped with $N_{\rm tx}=8$ antenna elements. All frequency-selective wireless fading channels were simulated as Rayleigh distributed considering $N_{\rm sub}=16$ SCs at ${\rm BW}=100$ MHz of bandwidth on the carrier frequency $f_c=2.4$ GHz.}
	\label{fig:Rates_vs_M}
\end{figure}

In Fig.~\ref{fig:Rates_vs_M}, the impact of the number $N_{\rm ris}$ of the RIS unit elements on the achievable sum rate is illustrated for the same designs as in Fig.~\ref{fig:Rates_vs_TX_Power}, considering the transmit power $P_{\max}=30$~dBm for every of the $K=2$ BSs with each equipped with $N_{\rm tx}=8$ antennas. Evidently, for all depicted curves, the sum-rate performance improves as $N_{\rm ris}$ increases, with the superior performance provided when BD-RISs are used. In contrast, for both designs based on conventional diagonal RISs  (i.e., $\vect{S}_1 = \vect{S}_2 = \mathbf{I}_{N_{\rm ris}}$), the achievable rates seem to saturate faster as $N_{\rm ris}$ becomes larger. This behavior showcases that BD-RISs can provide higher rates in cases of severe interference (the simulated scenario is operating at the high SINR regime). 
\begin{figure}[H]
	\centering
	\includegraphics[width=3.45in]{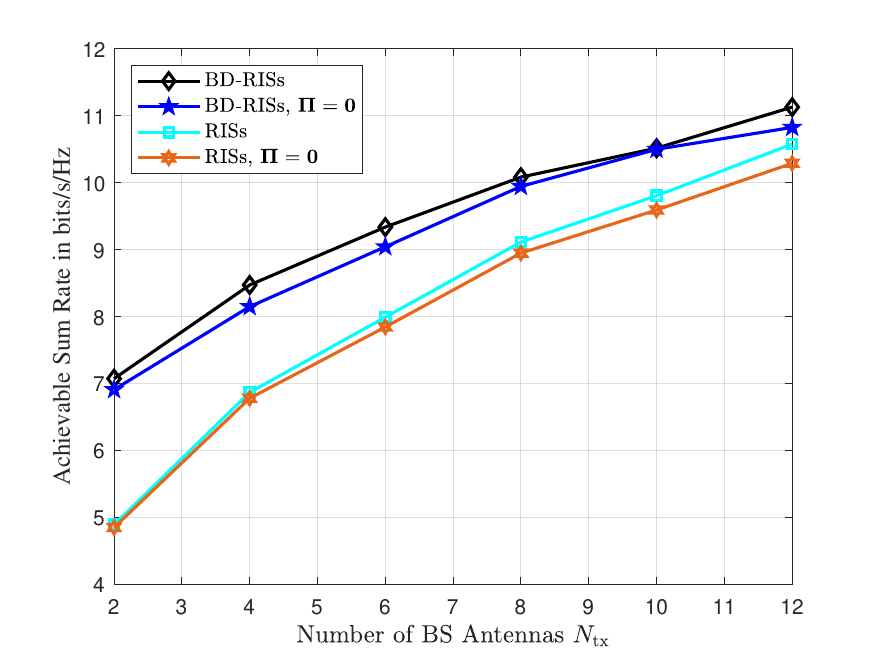}
	\caption{Achievable sum-rate performance in bits/s/Hz as a function of the common number $N_{\rm tx}$ of antenna elements at each of the $K=4$ BSs with common transmit power $P_{\max}=30$ dBm, considering $K=4$ frequency-selective BD-RISs or conventional diagonal RISs each with $N_{\rm ris}=100$ elements (i.e., $\vect{S}_k = \mathbf{I}_{100}$ $\forall k=1,2,3$, and $4$) for both cooperative and non-cooperative ($\vect{\Pi}=\vect{0}$) beamforming schemes using Algorithm~\ref{alg:OP_Overall_Distributed_Algorithm}. All frequency-selective wireless fading channels were simulated as Rayleigh distributed considering $N_{\rm sub}=16$ SCs at ${\rm BW}=100$ MHz of bandwidth on the carrier frequency $f_c=2.4$ GHz.}
	\label{fig:Rates_vs_N}
\end{figure}

Finally, in Fig.~\ref{fig:Rates_vs_N}, the achievable sum-rate performance is illustrated as a function of the common number $N_{\rm tx}$ antennas at each of the $K=4$ BSs with common transmit power $P_{\max}=30$ dBm, considering $K=4$ frequency-selective BD-RISs or conventional diagonal RISs each with $N_{\rm ris}=100$ elements (i.e., $\vect{S}_k = \mathbf{I}_{100}$ $\forall k=1,2,3$, and $4$) for both cooperative and non-cooperative ($\vect{\Pi}=\vect{0}$) beamforming schemes using Algorithm~\ref{alg:OP_Overall_Distributed_Algorithm}. It is shown that, for all schemes, the performance does not increase substantially with increasing $N_{\rm tx}$. Notably, the performance gains achieved by the cooperative schemes become substantially more pronounced than those of their non-cooperative counterparts, thereby underlining the effectiveness and advantage of the proposed distributed cooperative design framework. Furthermore, the results reveal that, as $N_{\rm tx}$ increases, the performance gap between the schemes employing BD-RISs and those using diagonal ones gradually diminishes, for both cooperative and non-cooperative schemes.

\section{Conclusions}\label{sec:conclusions}
This chapter highlighted the transformative role of RISs motivating the revolutionary concept of SWEs. Moving beyond the traditional view of the wireless channel as a random, uncontrollable medium, metasurfaces with dynamically reconfigurable EM responses have the potential to turn walls, ceilings, and objects in the wireless environment into programmable entities that can shape wave propagation in space, frequency, and even time. By embedding electronically tunable RISs into the environment, the radio medium itself becomes a design component that can be co-optimized with transceiver signal processing, rather than a constraint to be simply estimated and compensated.

The chapter overviewed the RIS operating principles and state-of-the-art hardware architectures, including passive, active, simultaneously reflecting and absorbing or transmitting, and BD-RIS designs, as well as the emerging XL MIMO architecture of DMAs. Key performance objectives and use cases of RIS-enabled SWEs, including SE and EE, EMF exposure reduction and sustainability, reliability, physical-layer security, energy harvesting, localization/sensing and ISAC, as well as the emerging paradigm of OTA (a.k.a. wave domain) computing, were discussed. Focusing on the recent trend of BD-RISs, two distributed designs of respective SWEs are presented. The first design dealt with a multi-UE MISO system operating within the area of influence of a SWE comprising multiple BD-RISs. A novel HDF ML framework based on MBACNNs, NN parameter sharing, and NE-based training was presented, which enables online mapping of channel realizations to the configurations of the multiple distributed BD-RISs as well as the BS transmit precoding matrix for the multi-UE data transmission. The presented performance evaluation results showcase that the distributedly optimized RIS-enabled SWE achieves near-optimal sum-rate performance with low online computational complexity. The second design was intended for a wideband interference MISO broadcast system, where each BS exclusively controls one BD-RIS to serve its assigned UE group. A cooperative optimization framework that jointly designs the BS transmit precoders as well as the tunable capacitances and switch matrices of all BD-RISs was presented. The provided numerical investigations verified the superior sum-rate performance of the designed RIS-enabled SWE for multi-cell MISO networks over benchmark schemes considering non-cooperative configuration and conventional diagonal metasurfaces.

\bibliographystyle{IEEEtran}
\addcontentsline{toc}{section}{References} 
\bibliography{references}

\end{document}